\pdfoutput=1 
\documentclass[11pt]{article}
\usepackage[margin=1in]{geometry}
\usepackage{amsthm}
\begingroup
\makeatletter
\@for\theoremstyle:=theorem,lemma,definition,claim,corollary\do{%
\expandafter\g@addto@macro\csname th@\theoremstyle\endcsname{%
\addtolength\thm@preskip\parskip
}%
}
\endgroup

\makeatletter
\def\thm@space@setup{%
  \thm@preskip=\parskip \thm@postskip=0pt
}
\makeatother

\newtheorem{theorem}{Theorem}
\newtheorem{definition}{Definition}
\newtheorem{lemma}[theorem]{Lemma}
\newtheorem{claim}[theorem]{Claim}

\usepackage{parskip}
\usepackage{graphicx}
\usepackage{amssymb}
\usepackage{amsmath}
\usepackage{epsfig}
\usepackage{subfig}
\usepackage{url}
\usepackage{enumerate}
\usepackage{tikz}
\usetikzlibrary{plotmarks}
\usetikzlibrary{calc}
\usepackage[hyperfootnotes=false]{hyperref}
\hypersetup{colorlinks=true, linkcolor=black, citecolor=black, urlcolor=black}
\newcommand{\mainfile}{\newif\ifmainfile\mainfilefalse}

\newcommand{\abs}[1]{\left|#1\right|}
\newcommand{\floor}[1]{\left\lfloor#1\right\rfloor}
\newcommand{\ceil}[1]{\left\lceil#1\right\rceil}
\newcommand{\norm}[2]{\left \lVert#2\right \rVert_{#1}}
\newcommand{\R}{{\mathbb R}}
\newcommand{\Hc}{{\cal H}}
\newcommand{\Gc}{{\cal G}}

\DeclareMathOperator{\E}{E}
\DeclareMathOperator{\err}{Err}

\makeatletter
\let\UrlSpecialsOld\UrlSpecials
\def\UrlSpecials{\UrlSpecialsOld\do\/{\Url@slash}\do\_{\Url@underscore}}%
\def\Url@slash{\@ifnextchar/{\kern-.11em\mathchar47\kern-.2em}%
    {\kern-.0em\mathchar47\kern-.08em\penalty\UrlBigBreakPenalty}}
\def\Url@underscore{\nfss@text{\leavevmode \kern.06em\vbox{\hrule\@width.3em}}}
\makeatother

{\makeatletter
 \gdef\xxxmark{%
   \expandafter\ifx\csname @mpargs\endcsname\relax 
     \expandafter\ifx\csname @captype\endcsname\relax 
       \marginpar{xxx}
     \else
       xxx 
     \fi
   \else
     xxx 
   \fi}
 \gdef\xxx{\@ifnextchar[\xxx@lab\xxx@nolab}
 \long\gdef\xxx@lab[#1]#2{{\bf [\xxxmark #2 ---{\sc #1}]}}
 \long\gdef\xxx@nolab#1{{\bf [\xxxmark #1]}}
 \long\gdef\xxx@lab[#1]#2{}\long\gdef\xxx@nolab#1{}%
}

\def\keywords#1{\vspace{7pt}\noindent{\it Keywords}\/:\ #1}

\begin{document}

\title{Compressive Sensing with Local Geometric Features}
\author{Rishi Gupta\\MIT CSAIL\and Piotr Indyk\\MIT CSAIL\and Eric Price\\MIT CSAIL\and Yaron Rachlin\\MIT Lincoln Laboratory}
\date{}
\maketitle

\begin{abstract}

We propose a framework for compressive sensing of images with \emph{local distinguishable objects}, such as stars, and apply it to solve a problem in celestial navigation.
Specifically, let $x \in \R^N$ be an $N$-pixel image, consisting of a small number of local distinguishable objects plus noise. Our goal is to design an $m\times N$ \emph{measurement matrix} $A$ with $m \ll N$, such that we can recover an approximation to $x$ from the measurements $Ax$.

We construct a matrix $A$ and recovery algorithm with the following properties: (i) if there are $k$ objects, the number of measurements $m$ is $O((k\log N)/(\log k))$, undercutting the best known bound of $O(k \log (N/k))$ (ii) the matrix $A$ is very sparse, which is important for hardware implementations of compressive sensing algorithms, and (iii) the recovery algorithm is empirically fast and runs in time polynomial in $k$ and $\log(N)$.

We also present a comprehensive study of the application of our algorithm to \emph{attitude determination}, or finding one's orientation in space. Spacecraft typically use cameras to acquire an image of the sky, and then identify stars in the image to compute their orientation. Taking pictures is very expensive for small spacecraft, since camera sensors use a lot of power. Our algorithm optically compresses the image before it reaches the camera's array of pixels, reducing the number of sensors that are required.

\keywords{Compressive sensing; Sparse matrices; Attitude determination; Star tracking.}

\end{abstract}

\section{Introduction}\label{sec:introduction}

\subsection{Compressive sensing}\label{sec:compressive-sensing}
Traditional approaches to image acquisition first capture an entire $N$-pixel image and then process it for compression, transmission, or storage. Often, the image is captured at very high fidelity, only to be immediately compressed after digitization. In contrast, compressive sensing uses prior knowledge about a signal to obtain a compressed representation directly, by acquiring a small number of nonadaptive linear measurements of the signal in hardware~\cite{CRT06:Stable-Signal,Don06:Compressed-Sensing}.
Formally, for an image represented by a vector $x$, we acquire the \emph{measurement vector}, or \emph{sketch}, $Ax$, where $A$ is an $m \times N$ matrix. The advantage of this architecture is that it uses fewer sensors, and therefore can be cheaper and use less energy than a conventional camera~\cite{DDTLTKB,FTF,Rom}.

In order to reconstruct the image $x$ from the lower-dimen\-sion\-al sketch $Ax$, we assume that the image $x$ is \emph{$k$-sparse} for some $k$ (i.e., has at most $k$ non-zero coordinates) or at least be well-approximated by a
$k$-sparse vector. Then, given $Ax$, one finds an approximation to
$x$ by performing {\em sparse recovery}.  The problem is
typically defined as follows: construct a matrix $A$ such that, for
any signal $x$, we can recover a vector $\hat{x}$ from $Ax$ satisfying
\begin{equation}
\label{e:lplq}
\norm{1}{x-\hat{x}} \le C \cdot \err^1_k(x),
\end{equation}
where $\err^1_k(x)=\min_{k\mbox{-sparse } x'}  \norm{1}{x-x'}$ and $C$ is the {\em approximation factor}.
Note that if $x$ is $k$-sparse, then $\err^1_k(x)=0$, and therefore $\hat{x}=x$.
Sparse recovery also has applications to other areas, such as data
stream computing~\cite{Muthu:survey,I-SSS}.

The problem of designing matrices $A$ and corresponding recovery algorithms has been a subject of extensive study over the last few years, with the goal of designing schemes that enjoy good compression rate (i.e., low values of $m$) as well as good algorithmic properties such as low encoding complexity and quick recovery times. Low encoding complexity is often achieved by using matrices that are \emph{binary} (entries chosen from $\{0,1\}$ or $\{-1,1\}$), or that have low \emph{column sparsity}. Column sparsity is the average number of non-zero entries per column, namely the average number of buckets into which each coordinate of the signal gets split.
 It is known by now that there exist binary
matrices $A$ and associated recovery algorithms that produce
approximations $\hat{x}$ satisfying Eq.~\ref{e:lplq} with constant approximation factor $C$ and sketch
length $m=O(k \log (N/k))$.
In particular, a
  random Bernoulli matrix~\cite{CRT06:Stable-Signal} or a random
  binary matrix with column sparsity $O(\log(N/k))$ \cite{BGIKS08} has
  this property with overwhelming probability.
It is also known that this sketch length is asymptotically optimal \cite{DIPW,FPRU}. See~\cite{GI} for an overview of compressive sensing using matrices with low column sparsity, along with~\cite{BI09} for a newer algorithm that we run experiments against (Section~\ref{sec:software}).

\subsection{Attitude determination}\label{sec:attitude}

Spacecraft determine their \emph{attitude}, or 3-axis orientation, by taking pictures of the sky ahead of them and identifying stars in the image. This function is encapsulated in a \emph{star tracker}, which is essentially a digital camera connected to a processor. To acquire the initial attitude, the camera
\begin{enumerate}
\item Takes a picture of the sky.
\item Identifies a set of starlike objects in the picture, and computes the centroid of each object.
\item Tries to match triangles and quadrilaterals formed by subsets of the centroids to an onboard database. A match provides approximate attitude information.
\item Uses the onboard database to determine a set of stars that it expects are in the picture, along with their approximate locations. Refines the attitude information by centroiding those stars as well.
\end{enumerate}
Most of the time, a star tracker has knowledge about its approximate attitude, either from the previous attitude computation or from other sensors. In that case, it goes from Step 1 directly to Step 4, in what we call \emph{tracking mode}.  \cite{Lie} has an overview of the process.

There are two types of sensors used in star tracker cameras, CCD (charge-coupled device), and CMOS (complimentary metal-oxide semiconductor). CCD sensors have low noise and capture a high fraction of the incoming signal, but have power usage and manufacturing costs that are super-linear in the number of pixels and that are high in absolute terms. On the other hand, CMOS sensors use little power, are cheap to manufacture, and allow random access to pixel values (important for tracking mode), but capture less of the incoming signal, and are very noisy. Most spacecraft use CCD cameras, but smaller or cheaper spacecraft use CMOS, taking a factor of 10 or higher hit in precision.

\subsection{Motivation for a new algorithm}

Ideally, we would like to have a camera with the precision of a high pixel CCD camera, but without the extra power and manufacturing costs that drive small spacecraft to use CMOS. The pictures taken by star trackers are naturally very sparse, in that most pixels are either empty or contain small stars not used in Steps 3 or 4. Also, the algorithms for Step 3 are very robust, and can tolerate a substantial fraction of bogus centroid information \cite{Mor}. A compressive sensing solution would optically compress the incoming picture, to reduce the size of the CCD array.

However, a standard compressive sensing solution runs into several major problems. First, the $L_1$ mass of the small stars is large compared to the $L_1$ mass of the signal. In other words, $\err_k^1(x) = O(\|x\|_1)$, and so Eq.~\ref{e:lplq} (Section~\ref{sec:compressive-sensing}) gives no guarantee at all. Second, the signal to noise ratio on star trackers is already low, and each non-zero entry of the sensing matrix $A$ will add substantial signal-independent noise to the final measurement.\cite{Holst07} Finally, each star is spread over multiple pixels, and makes only a small contribution to some of them. These pixels are needed to find the centroid of the star properly, but compressive sensing recovery techniques are only designed to recover the biggest pixels well.

We address many of these concerns by focusing on a compressive sensing algorithm where $A$ is very sparse. Compressive sensing algorithms with sparse measurement matrices are of general interest as well. Potential advantages in non-star tracking electronic compressive imagers include reduced interconnect complexity \cite{Meindl03}, low memory requirements for storing the measurement matrix, and gain in image acquisition speed due to reduced operations.

Unfortunately, it is known~\cite{Me} that any \emph{deterministic} scheme with guarantee as in Eq.~\ref{e:lplq} requires column sparsity of $\Omega(\log(N/k))$. In the randomized case, where $A$ is a random variable, and Eq.~\ref{e:lplq} is required to hold only with constant probability over the choice of $A$, the same paper shows that any binary matrix $A$ must have column sparsity as stated.

In this paper we overcome the above limitations by employing a two-fold approach. First, we consider a class of images where the $k$ large coefficients or $k$ local objects can be distinguished from each other. Second, we relax the recovery guarantee, by requiring that only a constant fraction of the objects are recovered correctly, and only with constant probability.


\subsection{Model description}\label{sec:model-description} Our model for sparse images is motivated by astronomical imaging, where an image contains a small number of \emph{distinguishable} objects (e.g., stars) plus some noise. We model each object as an image contained in a small $w \times w$ bounding box, for some $w = O(1)$. The image is constructed by placing $k$ objects in the image in an arbitrary fashion, subject to a \emph{minimum separation constraint}. The image is then modified by adding \emph{noise}. We formalize the notions of minimum separation constraint, distinguishability, and noise in the rest of this section. Some of the definitions below are illustrated in Appendix~\ref{sec:inpicture}.

Let $x$ be an $N$-dimensional real vector, and assume $N=n^2$ for an integer $n$. We will treat $x$ both as a vector and as an $n \times n$ matrix, with entries $x[i,j]$ for $i, j \in [n]$.  An {\em object} $o$ is a $w \times w$ real matrix. Let $\mathcal{O}=\{o_1, \ldots, o_k\}$ be a sequence of $k$ objects, and let $\mathcal{T}=\{t_1, \ldots, t_k\}$ be a sequence of translations in $x$, i.e., elements from $[n-w]^2$. We say that $\mathcal{T}$ is {\em valid} if for any $i \neq j$ the translations $t_i$ and $t_j$ do not {\em collide}, i.e., we have $\|t_i-t_j\|_{\infty} \ge w'$ for some constant separation parameter $w' = \Omega(w)$. For $o \in \mathcal{O}$ and $t=(t_x,t_y) \in \mathcal{T}$, we define $t(o)$ to be a $w \times w$ matrix indexed by $\{t_x, \ldots, t_x+w-1\} \times \{t_y, \ldots, t_y+w-1\}$. Using somewhat sloppy notation, the {\em ground truth image} is then defined as $x=\sum_i {t_i(o_i)}$.

During our algorithm, we impose a grid $G$ on the image with cells of size $w' \times w'$. Let $x_c$ be the image (i.e., an $w'^2$-dimensional vector) corresponding to cell $c$. We then use a \xxx{Does it have to be linear?}projection $F$ that maps each sub-image $x_c$ into a {\em feature vector} $F(x_c)$. If $y\subset x_c$ for some cell $c$ and some set of pixels $y$, we use $F(y)$ to denote $F(x_c)$ after the entries of $x_c\setminus y$ are set to 0. If $y$ is not a cell and not contained in a cell, we leave $F(y)$ undefined.

The distinguishability property we assume is that for any two distinct $o,o'$ from the objects $\mathcal{O} \cup \{\emptyset\}$, and for any two translations $t$ and $t'$, we have $\|F(t(o))-F(t'(o'))\|_\Gamma > T$ (when it is defined) for some threshold $T>0$ and some norm $\|\cdot\|_\Gamma$. In other words, different objects need to look different under $F$. For concreteness, the features we exploit in the experimental section are the {\em magnitude} (the sum of all pixels in the cell) and {\em centroid} (the sum of all pixels in the cell, weighted by pixel coordinates), since the magnitudes of stars follow a power law, and the centroid of a star can be resolved to $.15$ times the width of a pixel in each dimension (Section~\ref{sec:operation}). The distinguishability constraint is what ultimately allows us to undercut the usual lower bound by a factor of $\log k$.

The observed image $x'$ is equal to $x+\mu$, where $\mu$ is a noise vector. The threshold $T$ determines the total amount of noise that the algorithm tolerates. Specifically, let $\|\mu\|_F = \sum_c \|F(\mu_c)\|_\Gamma$, where $\mu_c$ is the noise corresponding to cell $c$. We assume that $\|\mu\|_F < \gamma kT$ for some small constant $\gamma>0$, and make no other assumptions about the noise.

\subsection{Results and techniques}

\subsubsection{Theoretical result} Assume sparsity parameter $k \ge
C\log N$ for some constant $C$, and prior knowledge of $k$ and the
distinguishability parameter $T$. We construct a distribution over
random binary $m \times N$ matrices $A$, such that given $Ax'$ for
$x'$ described above, we recover, with constant probability, a set $D$
of $k$ cells, such that at least $k/2$ of the cells fully containing
an object are included in $D$\footnote{From this, a simple min or
  median process can be used to recover an approximation to $x_c$ for
  any $c \in D$. See Section II.A of~\cite{GI} for an
  explanation of the technique.}.  The matrix has column sparsity
$O(\log_k N)$, and has $m=O(k \log_k N)$ rows. Note that we trade off
column sparsity and compression. If (say) $k=N^{1/2}$, then the column
sparsity is constant, and $m=O(N^{1/2})$.  The running time for the
recovery procedure is $O(k^3 \log^3_k n)$.

\subsubsection{Empirical result} We implement a standard attitude determination routine, with the picture acquisition step replaced with a simplified version of the theoretical algorithm. Our algorithm performs better recovery on small numbers of measurements and is orders of magnitude faster than comparable compressive sensing methods.

\subsubsection{Our techniques}  Our construction of the measurement matrix resembles those of other algorithms for sparse matrices, such as
Count-Sketch \cite{CCF} or Count-Min~\cite{CM03b}: we ``hash'' each cell $c$ into each of $s=O(\log_k N)$ arrays of $q=O(k)$ ``buckets'', and sum all the cells hashed to the same bucket. Each bucket defines one measurement of $w'^2$ pixels, which gives $m=O(k \log_k N)$.  Hashing is done using either the Reed-Solomon code or the Chinese Remainder code\footnote{Note that our use of the Chinese Remainder code does not incur any additional polylogarithmic factors.}.

The recovery process is based on the following novel approach. For simplicity, assume for now that the image contains no noise, and ignore the effect of two different objects being hashed to the same bucket. In this case, all buckets containing distinct objects are distinguishable from each other. Therefore, we can group non-empty buckets into $k$ clusters of size $s$, with each cluster containing buckets with a single value. Since $q^s > N$, each cluster of buckets uniquely determines the cell in $x$ containing the object in those buckets.

In order to make this approach practical, however, we need to make it robust to errors. The errors are due to distinct objects being hashed to the same bucket, the noise vector $\mu$, and the grid cutting objects into pieces. Because of these issues, the clustering procedure aims to find clusters containing elements that are close to each other, rather than equal, and the procedure allows for some small fraction of outliers~\cite{CKMN}. For this purpose, we use the approximation algorithm for the $k$-center problem with outliers, which correctly clusters a constant fraction of the buckets. To handle the buckets that are grouped incorrectly, we construct our hash function using a constant rate error-correcting code~\cite{Guru}.

\section{Theoretical Results}
\label{cha:theory}
A graphical representation of the algorithm is presented in Appendix~\ref{sec:inpicture}. Our scheme works by ``hashing'' each cell $c$ into $s = O(\log_k N)$ different arrays of size $O(k)$.  We can think of this as a mapping $g$ from
$[N]$ to $[O(k)]^s$.  As long as each character of the mapping is
approximately pairwise independent, then (in expectation) most of the
$k$ objects will be alone in most of the array locations they map to.
Our reconstruction algorithm clusters the values in the cells,
giving us a noisy version $y'$ of the true codeword $y = g(c)$ with a
constant fraction of errors.  We then efficiently decode from
$y'$ to $c$.

The first and second sections below establish families $\mathcal{G}$ from which we will draw mappings $g$. The third uses $\mathcal{G}$ to construct a distribution over measurement matrices $A$, and the fourth presents an associated recovery algorithm. The main result is stated in Theorem~\ref{thm:main-theory}.

\subsection{Definitions and preliminaries}

We need an efficient error correcting code that is also
approximately pairwise independent in each character. This section gives
 precise definitions of our requirements, and the next section gives two codes that achieve them.

\begin{definition}\label{def:hashfamily}
  A hash family $\Hc$ of functions $h \colon A \to B$ is
  \emph{pairwise-independent} if, for any $x_1, x_2 \in A$ and $y_1,
  y_2 \in B$ with $x_1 \neq x_2$, we have $\Pr_{h \in
    \Hc}[h(x_1) = y_1 \cap h(x_2) = y_2] =
  \frac{1}{\abs{B}^2}$.
\end{definition}

For any prime $P \geq N$, the function family $\Hc_P: ax + b \pmod{P}$
for $a, b \in [P]$ is pairwise independent when viewed as a set of
functions from $[N]$ to $[P]$.

In many of our applications the range $B$ is the product of $s$
``symbols'' $B_1 \times \dotsb \times B_s$.  For a function $f\colon A
\to B$ and $i \in [s]$, we use $f_i(x)$ to denote the $i$th coordinate
of $f$.  When $B$ is a product space, we will sometimes settle for a
weaker notion of pairwise independence.  Rather than requiring
pairwise independence for the whole range, we only require approximate
pairwise independence in each coordinate:

\begin{definition}
  Let $B = B_1 \times \dotsb \times B_s$.  A hash family $\Hc$ of
  functions $h \colon A \to B$ is \emph{coordinatewise
    $C$-pairwise-independent} if, for all $i \in [s]$, any $x_1 \neq
  x_2 \in A$, and all $y_1, y_2 \in B_i$, we have $\Pr_{h \in
    \Hc}[h_i(x_1) = y_1 \cap h_i(x_2) = y_2]
  \leq \frac{C}{\abs{B_i}^2}$.
\end{definition}

\begin{definition}
  Let $B = B_1 \times \dotsb \times B_s$. A function $f \colon A \to
  B$ is \emph{$C$-uniform} if, for all $i \in [s]$ and all $y \in
  B_i$, $\Pr_{x \in A}[f_i(x) = y] \leq \frac{C}{\abs{B_i}}$.
\end{definition}

For any function $f\colon B \to D$ and family $\Hc$ of functions
$h\colon A \to B$, we use $f \circ \Hc$ to denote the family of $A \to
D$ functions $\{g(x) := f(h(x)) \mid h \in \Hc\}$.

\begin{claim}\label{claim:c2uniform}
  If $\Hc$ is pairwise-independent and $f$ is $C$-uniform, then $f
  \circ \Hc$ is coordinatewise $C^2$-pairwise-independent.
\end{claim}
\begin{proof}
  Let $\Hc$ be a family of functions $A \to B$ and let $f\colon B \to D =
  D_1 \times \dotsb \times D_s$.  Then for any $i \in [s]$, any $x_1
  \neq x_2 \in A$, and all $y_1, y_2 \in D_i$ we have:
  \begin{align*}
  &  \Pr_{h \in \Hc}[f_i(h(x_1)) = y_1 \cap f_i(h(x_2)) = y_2]\\
    =& \sum_{z_1, z_2 \in B} \Pr_{h \in \Hc}[h(x_1) = z_1 \cap h(x_2) = z_2 \cap f_i(z_1) = y_1 \cap f_i(z_2) = y_2]\\
    =& \sum_{z_1, z_2 \in B} \frac{1}{\abs{B}^2}\Pr\,[f_i(z_1) = y_1 \cap f_i(z_2) = y_2]\\
    =& \Pr_{z_1, z_2 \in B} [f_i(z_1) = y_1 \cap f_i(z_2) = y_2]\\
    =& \Pr_{z_1 \in B} [f_i(z_1) = y_1] \Pr_{z_2 \in B}[f_i(z_2) = y_2]\\
    \leq& \frac{C^2}{\abs{B_i}^2}
  \end{align*}
  as desired.
\end{proof}

\begin{definition}
  We say that a function $f\colon A \to B$ for $B = B_1 \times \dotsb
  \times B_s$ is an \emph{error-correcting code of distance $d$} if,
  for any two distinct $x_1, x_2 \in A$, $f(x_1)$ and $f(x_2)$ differ
  in at least $d$ coordinates.
  We say that $f$ is \emph{efficiently decodable} if we have an
  algorithm $\widetilde{f}^{-1}$ running in $\log^{O(1)} |B|$ time with $\widetilde{f}^{-1}(y) =
  x$ for any $x \in A$ and $y \in B$ such that $f(x)$ and $y$ differ
  in fewer than $d/2$ coordinates.
\end{definition}

Recall the hash family $\Hc_P : ax + b \pmod P$ of functions $[N] \to
[P]$.

\begin{claim}\label{claim:fhefficient}
  If $f$ is an efficiently decodable error-correcting code with
  distance $d$, then so is $f \circ h$ for every $h \in \Hc_P$ with $a
  \neq P$.
\end{claim}
\begin{proof}
  Since $a \neq P$, there exists an $a^{-1}$ modulo $P$, and we can
  efficiently compute it.  Hence $h$ is injective, so $f \circ h$ is
  an error-correcting code of distance $d$.  Furthermore, $(f \circ
  h)^{-1}(x) = a^{-1} (f^{-1}(x) - b) \pmod{P}$ is efficiently computable.
\end{proof}

\begin{definition}
  We say that a family $\Gc$ of functions $g \colon A \to B_1\times
  \dotsb \times B_s$ is an $(N, s, d)_q$-independent code if $\Gc$
  is coordinatewise 4-pairwise independent, $q \leq \abs{B_i} \leq 2q$
  for all $i \in [s]$, $\abs{A} \geq N$, and with probability at least
  $1 - 1/N$ over $g \in \Gc$ we have that $g$ is efficiently decodable
  with distance $d$.
\end{definition}
Claims~\ref{claim:c2uniform} and~\ref{claim:fhefficient} give the
following lemma:
\begin{lemma}\label{lemma:creatingcodes}
  If $f\colon [P] \to B_1 \times \dotsb \times B_s$ is $2$-uniform and
  efficiently decodable with distance $d$, and $q \leq \abs{B_i} \leq 2q$ for
  all $i$, then $f \circ \Hc_P$ is a $(N, s, d)_q$-independent code.
\end{lemma}

We now show that $(N, s, d)_q$-independent codes have few
collisions in expectation.

\begin{lemma}\label{lemma:few-collisions}
  Suppose $g\colon A \to B_1 \times \dotsc \times B_s$ is drawn from a
  $(N, s, d)_q$-independent code.  Let $S, S' \subset A$.  Define
  the set of ``colliding'' symbols
  \[
  X = \{(a, i) \mid a\in S, i \in [s], \exists a' \in S' \text{
    s.t. } g_i(a) = g_i(a'), a \neq a'\}.
  \]

  With probability at least $7/8$, $\abs{X} \leq 32\abs{S}\abs{S'}s/q$.
\end{lemma}
\begin{proof}
  We observe that
  \begin{align*}
    \E[\abs{X}] &= \sum_{i \in [s]} \sum_{a \in S} \Pr[(a,i) \in X]\\
    &\leq  \sum_{i \in [s]} \sum_{a \in S} \sum_{\substack{a' \in S'\\a' \neq a}} \Pr[g_i(a) = g_i(a')]\\
    &=  \sum_{i \in [s]} \sum_{a \in S} \sum_{\substack{a' \in S'\\a' \neq a}} \sum_{z \in B_i} \Pr[g_i(a) = z \cap g_i(a') = z]\\
    &\leq  \sum_{i \in [s]} \sum_{a \in S} \sum_{\substack{a' \in S'\\a' \neq a}} \sum_{z \in B_i} \frac{4}{\abs{B_i}^2}\\
    &\leq s\abs{S}\abs{S'} 4/ q.
  \end{align*}
  Hence, by Markov's inequality, $\abs{X} \leq
  32\abs{S}\abs{S'}s/q$ with probability at least $7/8$.
\end{proof}

\subsection{Two code constructions}
\label{sec:code-constructions}

We explicitly give two $(N, s, s-r)_q$-independent codes.  Both are
achievable for any parameters with $2N < q^r$ and $s < q / \log q$
(and the first code allows any $s < q$).  We let $P$ be a prime in
$\{\frac{1}{2}q^r, \dotsc, q^r\}$.

\subsubsection{Reed-Solomon code}
  Let $q\ge s$. The Reed-Solomon code $f_{RS} \colon [q^r] \to [q]^s$ is defined for
  $f(x)$ by (i) interpreting $x$ as an element of $\mathbb{F}_q^r$, (ii) defining
  $\chi_x \in \mathbb{F}_q[\xi]$ to be the degree $r-1$ polynomial with
  coefficients corresponding to $x$, and (iii) outputting $f(x) =
  (\chi_x(1), \dotsc, \chi_x(s))$.  It is well known to have distance
  $s-r$ and to be efficiently decodable~\cite{Jus76}.

\begin{claim}
  Let $f: [P] \to [q]^s$ be the restriction of $f_{RS}$ to $[P]$. Then
  $f$ is $2$-uniform, so $\Gc_{RS} = f \circ \Hc_P$ is a $(N, s,
  s-r)_q$-independent code.
\end{claim}
\begin{proof}
  Basic facts about polynomials give that $f_{RS}$ is $1$-uniform.
  Since $P \geq q^r/2$, $f$ is $2$-uniform.
  Lemma~\ref{lemma:creatingcodes} then gives the result.
\end{proof}

\subsubsection{Chinese remainder theorem (CRT) code}
\label{sec:crt}
Let $p_1, \dotsc, p_s \in [q, 2q]$ be distinct primes; note that the asymptotic distribution of prime numbers
implies $q/\log q =\Omega(s)$.  Hence for any $x \in [N]$, any $r$ of
the residues mod $p_1, \dotsc, p_s$ uniquely identify $x$.  The CRT
code $f_{CRT}\colon [P] \to [p_1] \times \dotsc \times [p_s]$ is
defined by taking the residues modulo each prime.  It has distance
$s-r$ and is efficiently decodable~\cite{GRS99}.

\begin{claim}
  The CRT code $f_{CRT}$ is $2$-uniform.  Hence $\Gc_{CRT} = f_{CRT}
  \circ \Hc_P$ is a $(N, s, s-r)_q$-independent code.
\end{claim}
\begin{proof}
  Let $i \in [s]$.  The projection of $f_{CRT}(x)$ onto its $i$th
  coordinate is $x \bmod p_i$.  Hence over the domain $[P]$, the ratio
  between the likelihood of the most common and the least common
  values in the range is $\frac{\ceil{P / p_i}}{\floor{P / p_i}} \leq
  2$.  Thus $f_{CRT}$ is $2$-uniform, and
  Lemma~\ref{lemma:creatingcodes} gives the result.
\end{proof}

\newcommand{\erasure}{\perp}

\subsection{The measurement matrix}
\label{sec:measurement-matrix}
In this section we present the measurement matrix $A$. A graphical
representation of the measurement process is presented on the first
page of Appendix~\ref{sec:inpicture}. Let $\mathcal{O}=\{o_1, \dotsc,
o_k\}$ be a sequence of $k$ features, and let $\mathcal{T}=\{t_1,
\dotsc, t_k\}$ be a sequence of (non-colliding) translations in
$x$. Let $\mu$ be the noise vector, and let $x'$ be the noisy
image. Finally, let $\alpha,\beta,\delta, \eta>0$ be (small)
constants whose values will be determined in the course of the
analysis.

At the beginning, we impose a square grid $G$ with  $w' \times w'$ cells on the image $x'$, such that $w'=w/\alpha$. The grid is shifted by a vector $v$ chosen uniformly at random from $[w']^2$.
Let $S'$ be the set of cells that intersect or contain some object $t_i(o_i)$, and $S \subset S'$ be the set of cells that fully contain some object $t_i(o_i)$.
Observe that a fixed object is fully contained in some cell with probability $(1-w/w')^2 > 1-2\alpha$, since each axis of the grid intersects the object with probability $w/w'$. This implies that the expected number of cells in $S'-S$ is at most
$2 \alpha k$, and by Markov's inequality $|S'-S| \le 16\alpha k$ with probability $7/8$.
From now on, we will assume the latter event holds. Let $k'=|S'|$. We choose $\alpha>0$ such that $k' \le 2k$.

Our measurement matrix $A$ is defined by the following linear
mapping. Let $G$ denote the set of cells.  Let $g \colon G \to B = B_1
\times \dotsb \times B_s$ be drawn from a $(N, s,
4(3\delta+\beta)s)_q$-independent code (such as either $\Gc_{RS}$ or
$\Gc_{CRT}$).  Moreover, we require that $k/q \le \eta$; such a code
is achievable per Section~\ref{sec:code-constructions} with
$s=\Theta(\log_k N)$ as long as $k > C \log N$ for some constant $C$
(such that both $q^{(1-4(3\delta + \beta))s} > k^{s/2} > 2N$ and $s <
\log N / \log k \leq q / \log q$).  For each $i =1, \dotsc, s$, we define a
$\abs{B_i}$-dimensional vector $z^i$ whose entries are elements in
$\R^{w'^2}$, such that for any $j$
\[  z^i_j = \sum_{g_i(c)=j} x'_c. \]
That is, we ``hash'' all cells into $|B_i| \ge q$ buckets, and sum all cells
hashed to the same bucket. The measurement vector $z=Ax'$ is now equal
to a concatenation of vectors $z^1, \dotsc, z^s$.  Note that the
dimension of $z$ is equal to $m= w'^2 \sum \abs{B_i} = O(qs) = O(k
\log_k N)$.

\subsection{Recovery algorithm}
\label{sec:recovery}
A graphical representation of the recovery process is presented on the second page of Appendix~\ref{sec:inpicture}.
The recovery algorithm starts by identifying the buckets that likely contain the cells from $S$, and labels them consistently (i.e., two buckets containing cells from $S$ should receive the same label), allowing for a small fraction of errors. We then use the labels to identify the cells.

The algorithm runs as follows. For a set $X \subset [s] \times [2q]$ of
pairs of indices, let $F(X)$ denote $\{F(z^i_j) : (i,j) \in X \}$.

\begin{enumerate}
\item Identify $R = \{(i,j) : \|F(z^i_j)\|_\Gamma \ge T/2\}$ (that is, $R$ contains the ``heavy cells'' of the measurement vector $z$).
\item Partition $R$ into sets $R',R^1,\dots,R^k$ such that $|R'| \le \delta sk$, and such that for each $1 \le l \le k$ the diameter of $F(R^l)$ is at most $T/2$.

\item For each label $l=1, \dotsc, k$, create a vector $u^l \in B$ such that for each $i=1, \dotsc, s$, $u^l_i=j$ if $(i,j) \in R^l$ (if there are many such $j$, ties are broken arbitrarily), or $u_i^l=\erasure$ (an arbitrary erasure symbol) if no such $j$ exists.
\item For each label $l=1, \dotsc, k$ apply the decoding algorithm\footnote{Technically, we replace each $\erasure$ in $u^l$ with an arbitrary $j$ before running the decoding algorithm, since the decoding algorithms don't know about $\erasure$.} for $g$ to $u^l$, obtaining a (possibly invalid) decoded cell $d^l$.
\end{enumerate}

We analyze the algorithm by keeping track of the errors at each step.
\medskip

\noindent \emph{Step 1}\ \ For any cell $c \in S$ and $i =1, \dotsc, s$, we say that $i$ {\em preserves} $c$ if $\|F(z^i_{g_i(c)}) - F(x_{c}) \|_\Gamma \le T/24$ and $g_i(c') \ne g_i(c)$ for all other $c' \in S$. That is, there is no collision from the hashing process, and the total amount of distortion due to the noise $\mu$ is small. Let $P = \{(i,g_i(c)) : i \mbox{ preserves } c\}$. Note that $P\subset R$. We show that $P$ is large and that most of $R$ is in $P$.

\begin{lemma}
\label{l:beta}
With probability at least $7/8$, \[ |P| \ge (1-\beta)sk. \]
\end{lemma}

\begin{proof} Consider any pair $(c,i) \in S \times \{1, \dotsc, s\}$, and let $j=g_i(c)$. If $i$ does not preserve $c$, it must be because either (i) there is another cell $c' \in S'$, $c' \neq c$ such that $g_i(c')=j$, or because (ii) the total noise affecting $z^i_j$, equal to $F(\mu^i_j)\le \sum_{g_i(c)=j} F(\mu_c)$, has norm at least $T/24$.

By Lemma~\ref{lemma:few-collisions} with probability at least $7/8$
the number of pairs affected by (i) is at most $32k s |S'|/q$.  The
event (ii) is determined by the noise vector $\mu$. However, for each
$i$, there are at most $\frac{\sum_c \norm{\Gamma}{F(\mu_c)}}{T/24} = \frac{\|\mu\|_F }{T/24} < 24 \gamma k$
additional cells $c \in S$ that are not preserved under $i$ due to
this reason, where the latter inequality follows from the assumption that $\norm{F}{\mu} < \gamma k T$ (Section~\ref{sec:model-description}).

Altogether, the total number of pairs $(c,i)$ such that $c$ is not preserved by $i$ is at most

\[ 32sk\abs{S'}/q + 24 \gamma sk \leq [32 \eta (1+16\alpha) +24\gamma]  s k =\beta sk \]
for some small constant $\beta$, as desired.
\end{proof}

\begin{lemma}
\label{l:RiPi}
With probability at least $3/4$, \[ |R \setminus P| \le \delta sk. \]
\end{lemma}

\begin{proof}
Any element $(i,j)$ of $R\setminus P$ (``heavy but not preserved'') must belong to one of the following three categories:
\begin{enumerate}
\item $j=g_i(c)$ for $c \in S$ such that $c$ is not preserved by $i$. By the previous lemma, there are at most $\beta s k$ such pairs $(c,i)$ with probability at least $7/8$.
\item $j=g_i(c)$ for some cell $c \in S'\setminus S$. There are at most $16\alpha s k $ such pairs $(c,i)$, with probability at least $7/8$.
\item The vector $F(\mu^i_j)=\sum_{g_i(c)=j} F(\mu_c)$ has norm at least $T/2$. There are at most $2 \gamma s k $ such pairs $(i,j)$.
\end{enumerate}
This implies that with probability at least $3/4$ the total number of pairs $(i,j) \in R\setminus P$ is at most
\[ (\beta + 16\alpha  + 2 \gamma) sk = \delta sk \]
for some small constant $\delta$, as desired.
\end{proof}

\noindent \emph{Step 2}\ \ Observe that the elements of $F(P)$ can be
clustered into $k$ clusters of diameter $T/12$. Thus, by the previous
lemma, there is a $k$-clustering of all but $\delta sk$ elements of
$F(R)$ such that the diameter of each cluster is at most $T/12$. We
now apply a $6$-approximation algorithm for this problem, finding a
$k$-clustering of $F(R)$ such that the diameter of each cluster is at
most $T/2$. Such an approximation algorithm follows immediately from
the $3$-approximation algorithm for $k$-center with outliers
in~\cite{CKMN}, which gives a $k$-clustering with radius at most $T/4$
and hence diameter at most $T/2$.  \medskip

\noindent \emph{Step 3}\ \ Consider cells $c,c' \in S$ such that  $c$ is preserved by $i$ and $c'$ is preserved by $i'$.
If $F(z^i_{g_i(c)})$ and  $F(z^{i'}_{g_{i'}(c')})$ belong to the same cluster, then it must be the case that $c=c'$, since otherwise the distance between them would be at least $T - 2T/24>T/2$. In other words, for each $l$, if $u^l \subset P \cap R^l$ contains at least one element of $P$, then all the elements of $u^l$ are ``derived'' from the same cell.

\begin{lemma} With probability at least 3/4, $u^1,\dotsc, u^k$ contain a total of at most $2\delta sk$ errors and $(\delta+\beta)sk$ erasures ($i,l$ such that $u_i^l = \erasure$).
\end{lemma}
\begin{proof}
Let $R'' = R \setminus R' = R^1\cup\cdots\cup R^k$. Let $P' = P \cap R'$, and $P'' = P \cap R''$.

Note that $|P'| \le |R'| \le \delta sk$.
Each error in $u^1, \dotsc, u^k$ corresponds to a unique element of $R'' \setminus P''$, and we have
\[ |R'' \setminus P''| \le |R'' \setminus P| + |P \setminus P''| \le |R \setminus P| + |P'| \le \delta sk + \delta sk = 2\delta sk. \]
Additionally, $\{u^1, \dotsc, u^k\}$ contains at least $P''$ elements total, and so the number of erasures is at most $sk - |P''| = sk - |P| + |P'| \le \beta sk + \delta sk$, where we use $|P| \ge (1-\beta) sk$ from Lemma~\ref{l:beta}.
\end{proof}

\noindent \emph{Step 4}\ \ We can replace erasures by errors, and conclude that $u^1, \dotsc, u^k$ have a total of at most $(3\delta+\beta)sk$ errors. It follows that at least $k/2$ of them have at most $2(3\delta+\beta)s$ errors, and therefore can be decoded. Therefore, the set $D=\{d^1, \dotsc, d^k\}$ contains at least $k/2$ elements of $S$.

The running time of the recovery algorithm is dominated by Step 2, where we approximate $k$-median with outliers via the method in \cite{CKMN}. This takes $O((ks)^3) = O(k^3 \log^3_k n)$ time.

\begin{theorem}\label{thm:main-theory}
  Assume $k \ge C\log N$ for some constant $C$, a signal $x$ with $k$
  objects, and a noise vector $\mu$, all subject to the constraints
  delineated in the Model description of
  Section~\ref{sec:introduction}.  There is a distribution over random
  binary $m \times N$ matrices $A$, $m=O(k \log_k N)$, and an
  associated recovery algorithm with the following property.  Suppose
  that the algorithm is given $Ax'$ for $x'=x+\mu$. Then the algorithm
  recovers (with probability at least $3/4$) a set $D$ of $k$ cells,
  such that at least $k/2$ of the cells fully containing an object are
  included in $D$.  Moreover, the algorithm runs in $O(k^3 \log^3_k
  n)$ time and the matrix has column sparsity $O(\log_k
  N)$.
\end{theorem}

\section{Applications to Attitude Determination}
\label{sec:experimental}\label{cha:application}

Star trackers determine their attitude, or 3-axis orientation, by taking pictures of the sky and identifying stars in the image. We provide a detailed review of the current technology in the first section below. We then present a compressive sensing algorithm for attitude determination, along with a discussion about hardware implementation. Finally, we present results from a software simulation of the algorithm.

\subsection{Current star trackers}
\label{sec:back-star-trackers}
A star tracker is essentially a digital camera, called a \emph{star camera}, connected to a microprocessor. We describe various characteristics of the camera hardware and star identification algorithms.

\subsubsection{Numbers}
\label{sec:numbers}

We first provide some numbers from \cite{Lie} and \cite{RL} to give a sense of scale. As of 2001, a typical CCD star tracker consumes 5-15W of power\xxx{\cite{Lie}}. A small spacecraft uses 200W of power, and a minimal one uses less than 100W\xxx{cite{RL}}, so this can be a substantial amount. 

A high-end star tracker can resolve approximately the same set of stars that an unaided human can on a moonless night away from all light pollution. The number of stars in a star tracker's database varies from 58 to many thousands. The camera's field of view can vary from $2\times 2$ degrees to $30\times 30$ degrees, or anywhere from .01\% to 4\% of the sky. For comparison, the full moon is about .5 degrees across, and an adult fist held at arm's length is about 10 degrees across. A CCD camera can have up to a million pixels, and the accuracy of the final attitude is usually around .001 degrees (1 standard deviation), compared to .01 degrees for the next best sensors. The attitude is updated anywhere from 0.5 to 10 times a second.

\subsubsection{CCD and CMOS}
\label{sec:ccd-and-cmos}

CCD (charge-coupled device), and CMOS/APS (complimentary metal-oxide semiconductor/active pixel sensor) are the two different types of sensors used in star cameras. We abuse notation and use CCD/CMOS to refer to the sensor, an $n\times n$ array of sensors, the architecture of this array, and the star cameras they are a part of. Both CCD and CMOS turn incoming photons into electrons using the photoelectric effect, \emph{read} the electrons as a single voltage or charge, and then \emph{digitize} the charge with an analog to digital converter (ADC).

The CCD has a single ADC located in the corner of the pixel array. A CCD array is read as follows: each pixel repeatedly transfers its charge into a neighboring pixel, so that the charge from any given pixel eventually travels a taxicab path to the ADC. Charges from different pixels are never combined, so there are a total of $\Theta(n^3)$ charge transfers. Since the ADC only digitizes one pixel at a time, it also takes $\Theta(n^2)$ time to read the whole array. In addition, each \emph{charge transfer} leaves a fraction $\epsilon$ of the electrons behind, where $1-\epsilon$ equals the \emph{charge transfer efficiency}. The electrons in the farthest pixels undergo $\Theta(n)$ charge transfers, and in practice it is costly to achieve $\epsilon < 10^{-5}$, which puts a bound on the maximum size of a CCD array \cite{Hol,Fos}. Even if future technology were to allow a better charge transfer efficiency, it is worth noting that each charge transfer uses a large constant amount of power, and that the total number of charge transfers is super-linear in the number of pixels.

On the other hand, CMOS devices have an ADC built into every pixel. This solves all of the problems noted above, and adds another important feature: random access reading. In other words, we can choose to read and digitize only a subset of the pixels, and in practice, that is done, saving power and subsequent digital processing costs \cite{Lie}. However, the ADCs take up valuable real estate, and reduce the percentage of the chip that is available to collect photons. CMOS devices also generate substantially more noise than CCDs, further reducing the signal to noise ratio \cite{Litw}.

In practice, many consumer products such as cell phone cameras use CMOS, while scientific instruments use CCDs. Nevertheless, star trackers on small or low-power-budget satellites are starting to use CMOS, forgoing factors of 10 and higher in precision. We give the specification of a CCD tracker and a CMOS tracker in current (2011) production to illustrate the difference, as well as some of the numbers in highlighted in Section~\ref{sec:numbers}. The CT-602 Star Tracker has a CCD camera and is made by Ball Aerospace \& Technologies. It uses 8-9W of power, weighs 5.5kg, has 6000 stars in its database, an $8\times 8$ degree field of view, $512\times 512$ pixels, an attitude accuracy of .0008 degrees, and updates 10 times a second \cite{Ball}. Comtech AeroAstro's Miniature Star Tracker has a CMOS camera, uses $<2$W of power, weighs .4-.8 kg, has 1200 stars in its database, a $24\times 30$ degree field of view, and $1024\times 1280$ pixels, but has an accuracy of only .03 degrees and updates only 2 times a second~\cite{Comtech}.

\subsection{Star tracker operation}
\label{sec:operation}

We now study the process of attitude determination in more detail. As noted in Section~\ref{sec:attitude}, we often have an approximate attitude and go directly from Step 1 to Step 4 below.


\bigskip\noindent\emph{0. Celestial intuitions}\ \ %
Let the \emph{apparent mass} (hereafter, \emph{mass}) of a star be the number of photons from the star that reach our camera. The masses of the stars in the night sky follow a power-law distribution, with exponent $-1.17$ \cite{Lie}.

The stars are essentially infinitely far away, and can be treated as point sources of light. In particular, if a star camera were perfectly focused, the light from any given star would land on exactly 1 pixel, and we would not be able to resolve the centroid of any star at higher than pixel resolution. Star camera lenses are therefore intentionally out of focus, so that the light from a star lands on multiple pixels, which we can then average to get a centroid with sub-pixel resolution. The blurring can be mimicked by convolving the image with a Gaussian of radius .5 pixels\footnote{Technically, we want to use an Airy function, not a Gaussian. But for a blur of radius .5 pixels a Gaussian is a good approximation.} \cite{Lie}, after which most of a star's mass lands on 4-6 pixels. Note that this number is independent of the star's mass, the field of view, or the total number of pixels.

Additionally, all the stars that are visible to star trackers (or humans) are a part of the Milky Way galaxy, which is shaped like a disk. This means that the stars are not distributed uniformly over the sky. The density of stars varies by a factor of 4 \cite{LOA}; Figure~\ref{fig:star-locations} in Section~\ref{sec:software} gives some intuition. Most star trackers with a small field of view are not able to resolve the area perpendicular to the galaxy. 

\bigskip\noindent\emph{1. Image acquisition}\ \ %
The expected number of photons (i.e. mass) captured by a pixel is proportional to the area of the lens and the exposure time. Photons hit a pixel, and eject an electron with some constant probability via the photoelectric effect. The pixel is then read and digitized (Section~\ref{sec:ccd-and-cmos}).


Almost all the noise is introduced at this step. The signal-dependent portion of the noise, \emph{shot noise}, is due to random photon arrival times. This is best modelled by a Poisson distribution, which can be approximated as a Gaussian with a standard deviation equal to the square root of the expectation \cite{Hol}. For example, if a pixel expects to see 900 photons, it will see 900 photons with a standard deviation of $\sqrt{900} = 30$ photons.

There is also a large per-pixel component of the noise, which depends on the hardware, temperature, and other factors. See \cite{Hol} for a summary.



\bigskip\noindent\emph{2. Centroiding}\ \ %
We locate a set of stars $S$ in the image, by finding pixels with a value several standard deviations above the mean. We then centroid each star, by either taking the weighted mean of a few neighboring pixels, or by doing a more complicated Gaussian fitting. For a bright star, this can give a centroid resolution of .1 pixels in each dimension, even under moderate noise.

There are several errors that can occur here. If the threshold for being a star is low, we may get stars that are composed only of noise. If our catalog is incomplete (as it almost certainly will be), we may get stars in $S$ that aren't in the catalog. Finally, radiation or dust can occasionally cause a pixel to have an arbitrary value, in which case it might get classified as a star.


\bigskip\noindent\emph{3. Star identification}\ \ %
In this step, we match the stars in $S$ to an onboard database. If $S$ has zero or one stars, there is nothing to do, and we give up. 
Some star trackers may be able to make an identification when $|S|=2$, but we assume that $|S|\ge 3$. We explain the algorithm from one common star identification algorithm \cite{Mor}, since almost all of them follow the same outline~\cite{SM}.

\emph{Preprocessing}\ \ Depending on the size of the lens, the capabilities of the processor, and the expected signal to noise ratio, anywhere from a few dozen to tens of thousands of stars are selected from a catalog. The star tracker firmware is loaded with a data structure $D$ that stores all pairs of stars that can appear in the same picture, along with their distances.

\emph{In space}\ \ We check subsets of three stars from $S$ to see if they can form a valid triangle using edges from $D$. Once we find one, (and if $|S|\ge 4$), we try to find a fourth star from $S$ such that all $\binom{4}{2}$ distances are consistent with $D$, in which case we declare a match. We then use the 3-4 matched stars to determine an approximate attitude.

Even with tens of thousands of stars, it is extremely unlikely that we find a match that is consistent with $D$ that ends up being wrong. In other words, most quadrilaterals consistent with $D$ are far away from all other such quadrilaterals, and most quadrilaterals formed using erroneous stars match nothing.
 However, it is possible for there to be no noise in Steps 1-2 and still not have a match, due to catalog errors and omissions.


\bigskip\noindent\emph{4. Precise attitude computation}\ \ %
We use the approximate attitude information and the onboard database to obtain the rough location of other stars in the picture. The attitude is then refined using the centroids of those stars as well.

\bigskip Note that with a CMOS imager, if we start with an approximate attitude, we can actually skip most of Step 1 as well, and only read and digitize the handful of pixels under the stars we want to centroid.

\subsection{Specialization of the general algorithm}
\label{sec:specialization}

We specialize the theoretical algorithm presented in Section~\ref{cha:theory} to obtain a new algorithm for attitude determination, which we call Attitude Determination Under Analog Folding (ADUAF). The local geometric objects turn into stars, and the features we use are star centroid and mass. We use $f_{CRT}$ as our underlying error correcting code, rather than $f_{RS}$ (Section~\ref{sec:code-constructions}).

The biggest change from the theoretical algorithm to the practical algorithm is that we assume the stars are randomly rather than adversarially distributed. This means we no longer need to compose our error correcting code with the hash family $\Hc_p$. Additionally, we no longer need to shift the grid $G$ by the random vector $v$, as we do in Section~\ref{sec:measurement-matrix}. In fact, the notion of a grid is no longer needed at all, and we allow the decoded cell $d^l$ (Section~\ref{sec:recovery}) to be any $w' \times w'$ patch of the original image.

Given the physical cost of splitting the signal, the column sparsity $s$ is set to the smallest value our algorithm can tolerate, which is 2. This has the additional effect of turning the clustering process from Step~2 of Section~\ref{sec:recovery} into a much simpler bipartite matching process.

Finally, rather than just recovering the $w' \times w'$ patch $d^l$, we recover the centroid of $d^l$, or the centroid of the star on $d^l$. We then run the recovered centroids through the star identification process of Section~\ref{sec:operation}.
The full algorithm is presented below.

\subsubsection*{Measurements}
 If $p_1$ and $p_2$ are the primes used to construct the CRT code in Section~\ref{sec:crt}, we find primes $p'_i$ such that $p'_i \approx \sqrt{p_i}$. Note that we don't literally need $p'_1$ and $p'_2$ to be prime, as long as they are relatively prime, since the Chinese Remainder Theorem from Step~4 of the recovery algorithm applies for relatively prime numbers as well. We will use the word \emph{prime} to refer to $p'_1$ and $p'_2$ even when they are just relatively prime.

\begin{figure}[h!]
\centering
\includegraphics{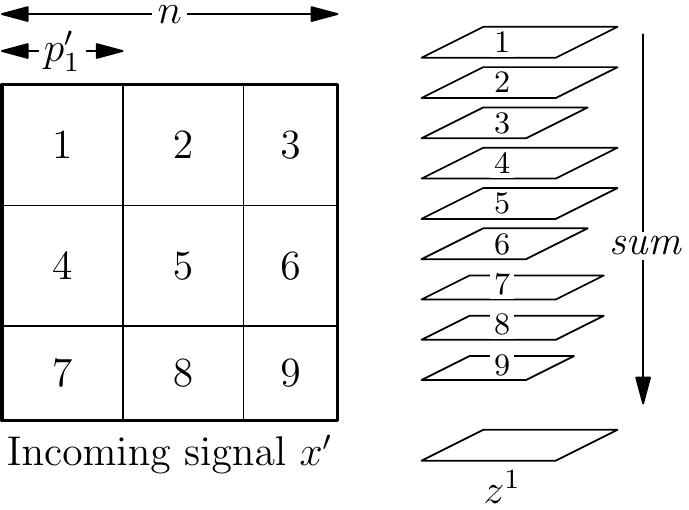} \hspace{.65in}
\includegraphics{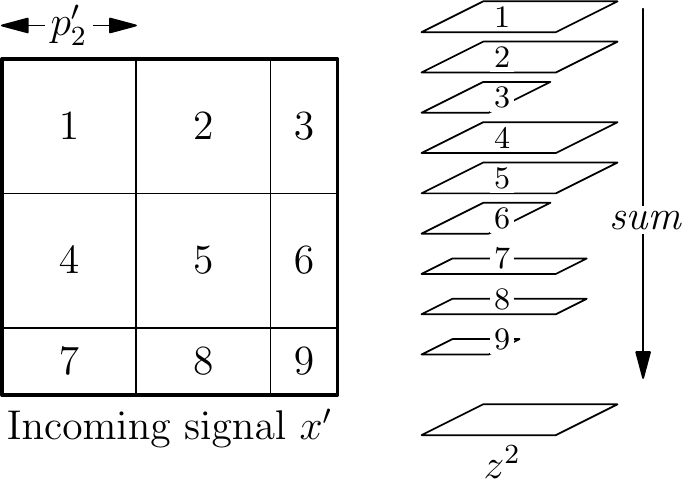}
\caption{\label{fig:folding} Taking measurements with column sparsity 2. Picture idea from \cite{UGNKJKB}.}
\end{figure}

Thinking of the noisy incoming signal $x'$ as an $n$-by-$n$ image, we define $z^i$ to be a $p'_i$-by-$p'_i$ image with
\[ z^i[j_1,j_2] = \sum_{\substack{c_1 \equiv j_1 \pmod{p'_i} \\ c_2 \equiv j_2 \pmod{p'_i}}} x'[c_1,c_2]. \]
We say we \emph{fold} $x'$ to obtain $z^i$. Figure~\ref{fig:folding} provides some intuition.

We actually think of each $z^i$ as a torus, and identify the top/bottom and left/right edges together, allowing recovery of stars that are on the edge of the folded picture. Since the stars were assumed to be randomly distributed in $x'$, they are randomly distributed within each $z^i$ as well. Also, one could define the ``measurement vector'' $z$ to be a 1-dimensional representation of the pair $(z^1,z^2)$, and construct a measurement matrix $A$ accordingly. However, it is more useful to think of each $z^i$ as a 2-dimensional image.

\subsubsection*{Recovery algorithm}
We follow the presentation of the corresponding paragraph in Section~\ref{sec:recovery}. For concreteness, we use a few specific numbers from our software implementation in Section~\ref{sec:software}.
\begin{enumerate}
\item In each $z^i$, we identify ten $3\times 3$ ($w\times w$) cells with high mass such that no two cells collide in more than four pixels. We allow the $3\times 3$ cells to wrap around the edges of $z^i$, as in a torus.
\item In place of the $k$-centering algorithm, we greedily choose up to eight pairs of cells $(c^1,c^2)$ from $(z^1,z^2)$ such that the feature vectors $F(c^1)$ and $F(c^2)$ are close. In our case, $F(c)$ is a triple of values: the mass of $c$, along with two coordinates for the centroid of $c$.
\item No longer needed. Each pair from Step~2 corresponds to one of the $u^l$.
\item In place of the error correcting code, we simply apply the Chinese Remainder Theorem in each dimension to recover a $3\times 3$ region of the original image for each pair from Step 2.
\end{enumerate}

We compute centroids of the recovered $3\times 3$ regions, by averaging the centroids (weighted by mass) of the corresponding $3\times 3$ regions of $z^1$ and $z^2$. This is now $S$ in Step 3 of Section~\ref{sec:operation}.

Step 1 above takes time linear in the number of measurements, or time $p_1+p_2\approx n$. Steps 2-4 take time at most quadratic in the number of cells chosen in Step 1 (ten, in our example).

\subsubsection*{Potential hardware}

We have identified several existing hardware architectures that implement the folding mechanism above. In our case, we would use $s$ copies ($s=2$ above) of the folding hardware, where $s$ is the column sparsity. We call each of the stacked squares in Figure~\ref{fig:folding} a \emph{piece} of the image. We denote the number of pieces by $\tau$ ($\tau=9$ in both $z^1$ and $z^2$ in Figure~\ref{fig:folding}). After being folded, each piece of the image lands on the \emph{focal plane}, which has an array of $z^i$ pixels.

The most plausible architecture uses mirrors to directly reflect the pieces of the image onto the focal plane. The single-pixel camera~\cite{DDTLTKB} implements a generalized version of this for $z^i=1$ using $n$ moving mirrors. We would use $\tau$ rigid mirrors for each $z^i$, which is a substantial simplification of what they implemented.

Another possibility is to use $\tau$ different lenses, each of which focuses one piece of the image onto the focal plane \cite{UGNKJKB}. The challenge with this design will be making all the lenses face the same direction.

\begin{figure}[h!]
\centering
\subfloat[From \cite{UGNKJKB}.]{\includegraphics{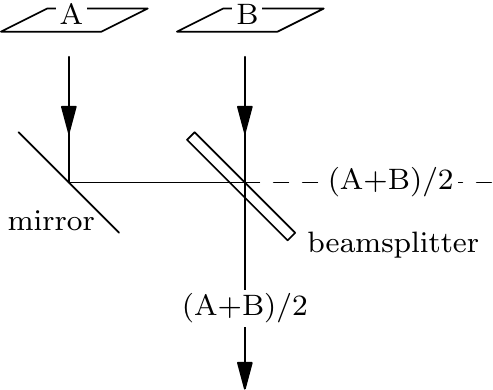}} \hspace{1in}
\subfloat[From \cite{TAN}.]{\includegraphics{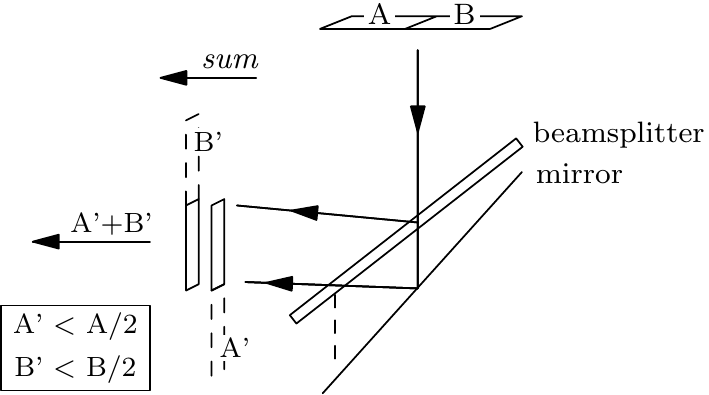}}
\caption{\label{fig:beamsplitters} Two examples of how to combine pieces of an image using a mirror and beamsplitter. Dashed lines indicate lost signal.}
\end{figure}

The remaining two architectures are designed for unrelated earth-based tasks, and are more wasteful of the signal than is appropriate for a star tracker. Nevertheless, they show the extent to which researchers are already thinking about hardware designs for folding.
Both of the designs use beamsplitters and mirrors to combine the signal, and lose a factor of $\tau$ from the signal in the process. Figure~\ref{fig:beamsplitters} depicts an element from each design. To extend \ref{fig:beamsplitters}(a) to $\tau$ pieces one would use $\log{\tau}$ different such image combiners, and to extend \ref{fig:beamsplitters}(b) to $\tau$ pieces one would use $\sqrt{\tau}-1$ stacked beamsplitters plus a mirror to fold in one direction, and $\sqrt{\tau}-1$ stacked beamsplitters plus a mirror to fold in the other.

Finally, it is possible we could split the signal and fold it after it hits the focal plane but before it gets digitized, which would not save on sensor costs, but would reduce the number of ADC computations, and reduce the load on the onboard processors. CCD images query an entire row of the image before digitizing \cite{Litw}, so folding in at least the first dimension could be relatively easy. CMOS imagers are already built directly onto integrated circuits, so additional circuitry for folding is likely to be cheap \cite{Robucci10}. Several (non-satellite) CMOS imagers have already been built that use dense compressive sensing matrices to reduce the number of measurements that need to be digitized \cite{Robucci10,Majidzadeh10}.


\subsection{Software implementation}
\label{sec:software}

We implement the algorithm presented in Section~\ref{sec:specialization}, and run experiments on simula\-ted images. Source code is available at \url{http://web.mit.edu/rishig/papers/local-geo/}.

\begin{figure}[h!]
\centering
\includegraphics[width=5in]{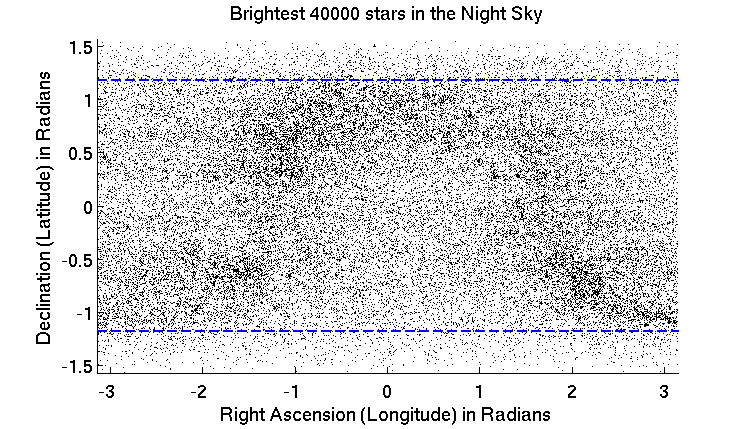}
\caption{\label{fig:star-locations} Mercator projection of the night sky. The dense omega-shaped region is the central disk of the galaxy. We test our algorithm on the area between the dashed blue lines.}
\end{figure}

For simplicity, we project the stars onto a rectangular interval, rather than operating on a sphere (Figure~\ref{fig:star-locations}). We index the rectangle by right ascension (longitude) $\alpha \in [-\pi,\pi]$ and declination (latitude) $\delta \in [-\pi/2,\pi/2]$; for example, $\delta = 0$ is the equator, and $\delta = \pi/2$ is the north pole. So that the approximation makes sense, we ignore the portion of the sky where $|\delta| > \pi/2 - \pi/8$ (the dashed blue lines in Figure~\ref{fig:star-locations}). This also has the effect of removing the portion of the sky that has the fewest stars, which some star trackers don't operate on anyway. We assume without loss of generality that the camera is axis-aligned.

We fix $n = 800$ ($N = 640000$) for all experiments. We expose the camera to a .08 radian by .08 radian (4.6 by 4.6 degree) patch of the sky, which means that a typical image will have 3 to 10 bright stars (10\%-ile to 90\%-ile) and 50 to 150 stars that act as background noise. Above, we defined the mass (apparent mass) of a star to be the number of photons from the star that hit our camera. In our pictures, if the median total star mass is scaled to 1, the 10\%-ile mass is .6, and the 90\%-ile mass is 2.25. These numbers were determined empirically from the Smithsonian Astrophysical Observatory (SAO) Star Catalog~\cite{SAO}. Recall the mass of the $j^{th}$ brightest star in the sky is $\Theta(j^{-1.17})$~\cite{Lie}.

\def\SAO{{\sc sao}}
We choose the stars for the preprocessed database $D$ (Step~3 of Section~\ref{sec:operation}) as follows. We extract a subset SAO$'$ from the full SAO Catalog, by taking the 10 most massive stars in every ball of radius .08 radians. SAO$'$ has 17100 stars, compared to 259000 stars in the SAO catalog.

To generate test images, we randomly select axis-aligned patches of the sky from the area between the blue lines of Figure~\ref{fig:star-locations}, and use the SAO star catalog to simulate star positions and mass. We convolve the stars with a Gaussian of radius $.5$ pixels, which causes the vast majority of a star's mass to fall within a $3\times 3$ box. We then apply Poisson noise to account for photon arrival times, and fold the image using two relatively prime numbers. We add Gaussian noise (amount varying by experiment) to each pixel of both folded images to account for all signal-independent sources of noise. To keep the recovery process simple, we do not account for the Poisson noise introduced by splitting the signal; in other words, we apply Poisson noise once to each star, whereas in a hardware implementation it is likely that Poisson noise will be applied independently to each folded copy of a star. \looseness=-1





\begin{figure}[h!]
\centering
\subfloat[Underlying noiseless signal $x$ (zoom in).]{\includegraphics[width=.4\textwidth, height=.36\textwidth]{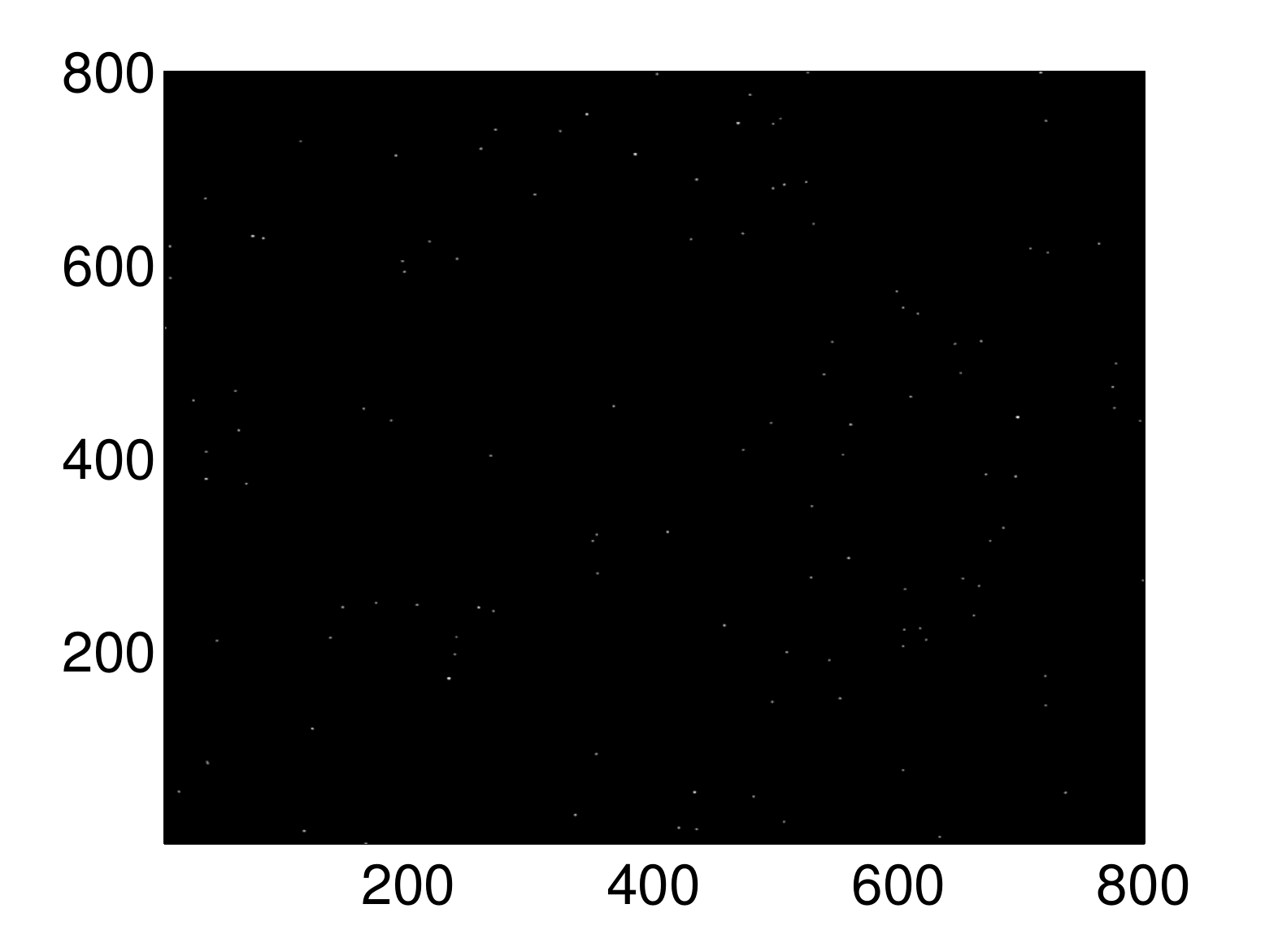}} \\
\subfloat[$p'= 29$, No Noise.]{\includegraphics[width=.37\textwidth, height=.32\textwidth]{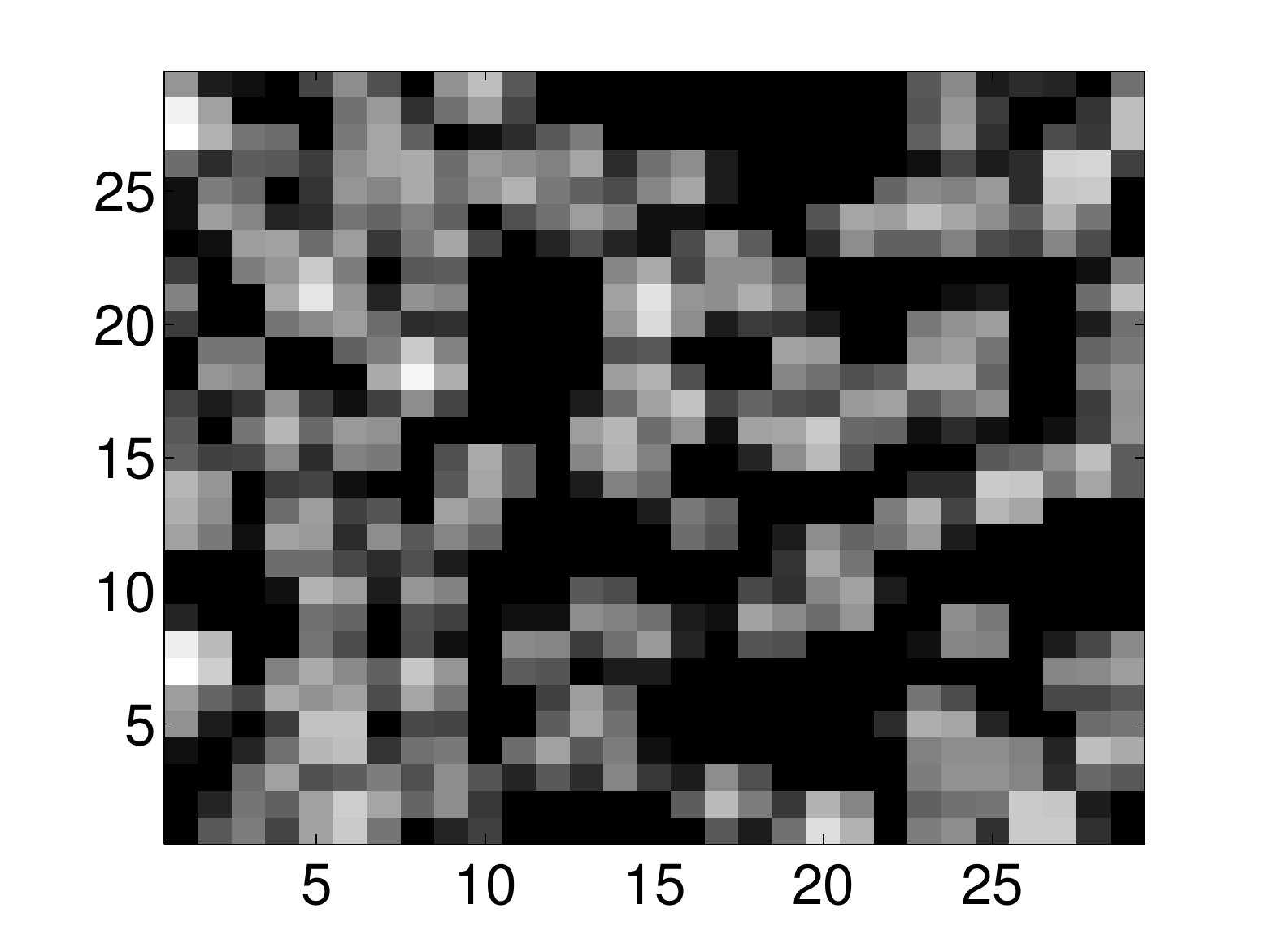}}
\subfloat[$p' = 29$, Noise = 150]{\includegraphics[width=.425\textwidth]{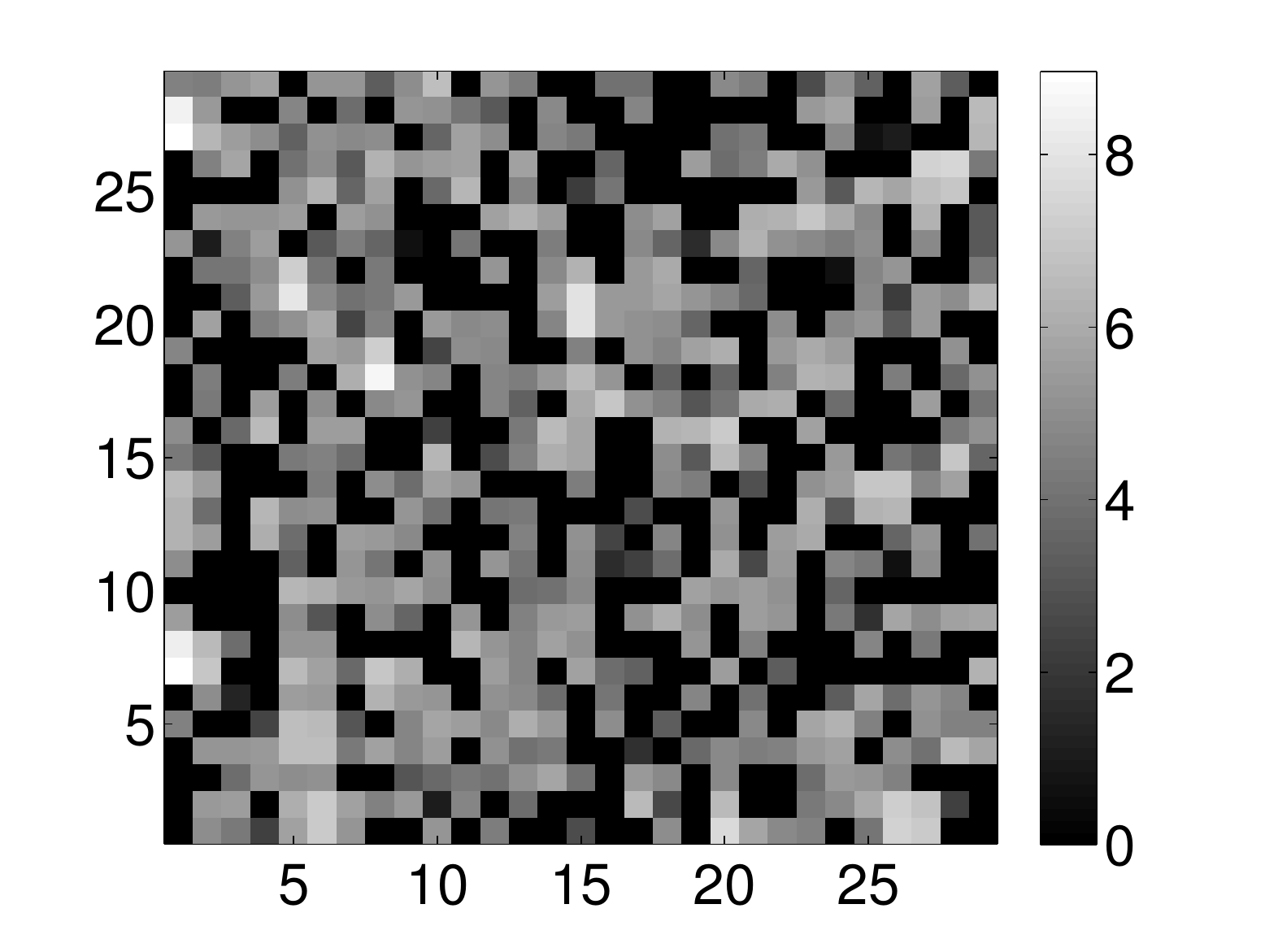}}
\caption{\label{fig:sample-pics} log(mass) in sample images from a representative part of the sky. The legend on the right applies to all three images.  We cut off pixels with values below 0 before taking the log in (c).}
\end{figure}

Figure~\ref{fig:sample-pics} has a few pictures to help build intuition about star density and what stars look like.
It is generally possible to see where the biggest stars are located, though some fraction of them get occluded by small stars or noise.

We run our algorithm, ADUAF, as well as Sequential Sparse Matching Pursuit (SSMP) on 159 images of the sky. SSMP includes a linear (in $N$) time sparse binary compressive sensing recovery algorithm, with recovery quality on par with other known methods~\cite{BI09}. We use the same folded images, or equivalently, the same measurement matrix, for both algorithms.

ADUAF recovers a list of centroids, and we find and centroid stars in the image recovered from SSMP. We run the star identification algorithm from Step~3 of Section~\ref{sec:operation} on the outputs of both of them, and declare success if they identify the stars correctly. We report our results as a function of the standard deviation of the added Gaussian noise (Figure~\ref{fig:pattern-matching}).


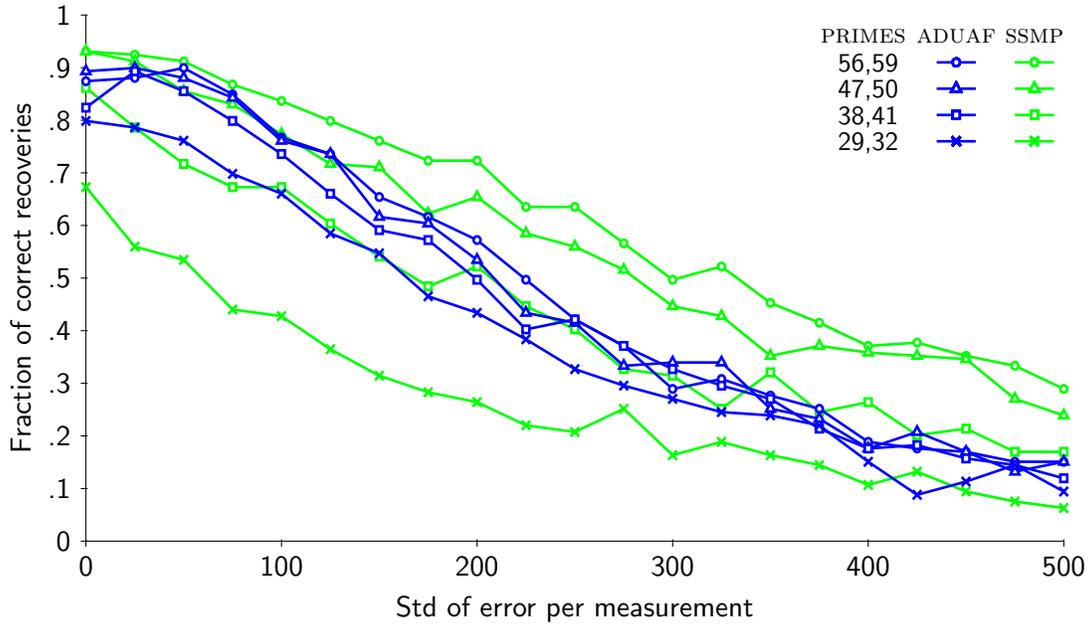
\begin{figure}[h!]
\centering
\providecommand{\mainfile}{\newif\ifmainfile\mainfiletrue}
\mainfile

\ifmainfile
\documentclass{article}
\usepackage{tikz}
\usetikzlibrary{plotmarks}
\usetikzlibrary{calc}
\begin{document}
\fi

\begin{tikzpicture}[font=\sffamily, xscale=.65, yscale=.7]
	\draw (0,0) -- coordinate (x axis mid) (20,0);
    	\draw (0,0) -- coordinate (y axis mid) (0,10);
    	\foreach \x / \r in {0/0,4/100,8/200,12/300,16/400,20/500} {
     		\draw (\x,1pt) -- (\x,-3pt)
			node[anchor=north] {\r};
        }
    	\foreach \y/\r in {0/0,1/.1,2/.2,3/.3,4/.4,5/.5,6/.6,7/.7,8/.8,9/.9,10/1} {
     		\draw (1pt,\y) -- (-3pt,\y)
     			node[anchor=east] {\r};
        }
	\node[below=0.6cm] at (x axis mid) {Std of error per measurement};
	\node[rotate=90, above=0.6cm] at (y axis mid) {Fraction of correct recoveries};

        \tikzset{ssmp/.style={green, line width=1pt}}
        \tikzset{srgf/.style={blue, line width=1pt}}

	\draw[ssmp] plot[thick, mark=*, mark options={fill=white}] file {graphics/ssmp.1};
	\draw[ssmp] plot[mark=triangle*, mark options={fill=white}, mark size=3] file {graphics/ssmp.2};
	\draw[ssmp] plot[mark=square*, mark options={fill=white}] file {graphics/ssmp.3};
	\draw[ssmp] plot[mark=x, mark options={fill=white}, mark size=3.5] file {graphics/ssmp.4};

	\draw[srgf] plot[thick, mark=*, mark options={fill=white}] file {graphics/srgf.1};
	\draw[srgf] plot[mark=triangle*, mark options={fill=white}, mark size=3] file {graphics/srgf.2};
	\draw[srgf] plot[mark=square*, mark options={fill=white}] file {graphics/srgf.3};
	\draw[srgf] plot[mark=x, mark options={fill=white}, mark size=3.5] file {graphics/srgf.4};

	\begin{scope}[shift={(17,9.6)}]

        \node at (-1.1,0) {\small \sc primes};
        \node at (-1,-.54) {\small 56,59};
        \node at (-1,-1.04) {\small 47,50};
        \node at (-1,-1.54) {\small 38,41};
        \node at (-1,-2.04) {\small 29,32};

     \begin{scope}[shift={(.4,0)}]
        \node at (.4,0) {\small \sc aduaf};
	\draw[srgf,yshift=-.5cm] (0,0) --
		plot[mark=*, mark options={fill=white}] (0.4,0) -- (.8,0);
	\draw[srgf,yshift=-1cm] (0,0) --
		plot[mark=triangle*, mark size=3, mark options={fill=white}] (0.4,0) -- (.8,0);
	\draw[srgf,yshift=-1.5cm] (0,0) --
		plot[mark=square*, mark options={fill=white}] (0.4,0) -- (.8,0);
	\draw[srgf,yshift=-2cm] (0,0) --
		plot[mark=x, mark size=3.5, mark options={fill=black}] (0.4,0) -- (0.8,0);
      \end{scope}

     \begin{scope}[shift={(2,0)}]
          \node at (.4,0) {\small \sc ssmp};
	  \draw[ssmp,yshift=-.5cm] (0,0) --
		plot[mark=*, mark options={fill=white}] (0.4,0) -- (0.8,0);
	\draw[ssmp,yshift=-1cm] (0,0) --
		plot[mark=triangle*, mark size=3, mark options={fill=white}] (0.4,0) -- (0.8,0);
	\draw[ssmp,yshift=-1.5cm] (0,0) --
		plot[mark=square*, mark options={fill=white}] (0.4,0) -- (0.8,0);
	\draw[ssmp,yshift=-2cm] (0,0) --
		plot[mark=x, mark size=3.5, mark options={fill=black}] (0.4,0) -- (0.8,0);
	\end{scope}

	\end{scope}

\end{tikzpicture}

\ifmainfile
\end{document}
\fi
\caption{\label{fig:pattern-matching} Experimental results. Each point on the figure is computed using the same 159 underlying images.}
\end{figure}

The first observation we make is that ADUAF works very well down to an almost minimal number of measurements. The product $p'_1p'_2$ has to be greater than 800, and the minimal set of primes is 26 and 31. As the number of measurements increases, SSMP catches up and surpasses ADUAF, but we note that running SSMP (implemented in C) takes 2.4 seconds per trial on a 2.3 GHz laptop, while ADUAF (implemented in Octave/Matlab) takes .03 seconds per trial. Computation power on a satellite is substantially lower than that of a low end laptop, and given that the entire acquisition has to happen in .1 to 2 seconds, it seems unlikely that any algorithm linear or near linear in $N$ is going to be practical. Finally, we note that the plot lines for both SSMP and ADUAF could be improved by a more sophisticated implementation of the star identification algorithm.

\section*{Acknowledgements}
The authors would like to thank Tye Brady and Ben Lane from Draper Laboratory for numerous conversations and for help with the data. We would also like to thank the anonymous reviewers for their thorough reviews and for helping clarify the presentation.

This research has been supported in part by a David and
Lucille Packard Fellowship, MADALGO (Center for Massive
Data Algorithmics, funded by the Danish National Research
Association) and NSF grant CCF-0728645. R.\ Gupta has been supported in part by a Draper Laboratory Fellowship. E.\ Price has been supported in part by an NSF Graduate Research Fellowship.

\appendix
\providecommand{\mainfile}{\newif\ifmainfile\mainfiletrue}
\mainfile

\ifmainfile
  \documentclass[11pt]{article}
  \usepackage{tikz}
  \usepackage[margin=1in]{geometry}
  \usetikzlibrary{calc}
  \pagestyle{empty}
  \begin{document}
\else
\fi

\onecolumn
\newif\ifpageone\pageonetrue
\newif\ifpagetwo\pagetwotrue

\newif\ifstarbw\starbwfalse
\newif\ifheavyonly\heavyonlyfalse

\def\starrad{.2}
\def\starcolorbw{black!50}
\newcommand{\stara}[2]{
  \begin{scope}[shift={(#1,#2)}]
  \ifstarbw
    \filldraw[\starcolorbw] (10:\starrad) \foreach \x in {10,130,...,360} {
      -- (\x:\starrad)
    } -- cycle;
  \else
    \filldraw[red!40!gray] (10:\starrad) \foreach \x in {10,130,...,360} {
      -- (\x:\starrad)
    } -- cycle;
  \fi
  \end{scope}
}
\newcommand{\starb}[2]{
  \begin{scope}[shift={(#1,#2)}]
  \ifstarbw
    \filldraw[\starcolorbw] (20:\starrad) \foreach \x in {20,110,...,360} {
      -- (\x:\starrad)
    } -- cycle;
  \else
    \filldraw[blue!50!gray] (20:\starrad) \foreach \x in {20,110,...,360} {
      -- (\x:\starrad)
    } -- cycle;
  \fi
  \end{scope}
}
\newcommand{\starc}[2]{
  \begin{scope}[shift={(#1,#2)}]
  \ifstarbw
    \filldraw[\starcolorbw] (20:\starrad) \foreach \x in {20,92,...,360} {
      -- (\x:\starrad)
    } -- cycle;
  \else
    \filldraw[green!40!gray] (20:\starrad) \foreach \x in {20,92,...,360} {
      -- (\x:\starrad)
    } -- cycle;
  \fi
  \end{scope}
}
\newcommand\noise[1]{
  \fill [black!40] #1 circle (2pt) ;
}
\def\mynode#{\vtop \bgroup \hsize 0pt \parindent 0pt 
        \rightskip = 0pt minus \maxdimen \let\next=}
\def\nn{4} 

\newcommand\fold[2]{
  \ifheavyonly \else
    \draw[step=1,gray,semithick] (0,0) grid (#1,1);
    \draw[step=1,gray,very thin] (#1,0) grid (#2,1);
  \fi
  \pgfmathmod{17.4}{#1}
  \stara{\pgfmathresult}{.6};
  \pgfmathmod{5.35}{#1}
  \starb{\pgfmathresult}{.5};
  \pgfmathmod{8.5}{#1}
  \starc{\pgfmathresult}{.43};
  \foreach \x / \y in \noisesmod {
    \pgfmathmod{\x}{#1};
    \fill[black!40] (\pgfmathresult,\y) circle (2pt);
  }
}

\tikzset{rectselect/.style={step=1,very thick}}
\tikzset{r1/.style={very thick, magenta!80!black}}
\tikzset{r2/.style={very thick, green!40!yellow!70!black}}
\tikzset{r3/.style={very thick, cyan!70!black}}

\newcommand\measurements{
  \begin{scope}[yshift=4.05cm]
    \ifheavyonly \clip (1,0) rectangle (4,1); \fi        
    \fold{5}{9};
  \end{scope}
  \begin{scope}[yshift=2.7cm]
    \ifheavyonly \clip (0,0)--(2,0)--(2,1)--(3,1)--(3,0)--(6,0)--(6,1)--(0,1)--cycle; \fi        
    \fold{7}{9};
  \end{scope}
  \begin{scope}[yshift=1.35cm]
    \ifheavyonly \clip (0,0)--(3,0)--(3,1)--(5,1)--(5,0)--(6,0)--(6,1)--(0,1)--cycle; \fi        
    \fold{8}{9};
  \end{scope}
  \begin{scope}[yshift=0cm]
    \ifheavyonly \clip (2,0)--(3,0)--(3,1)--(4,1)--(4,0)--(6,0)--(6,1)--(8,1)--(8,0)--(9,0)--(9,1)--(2,1)--cycle; \fi        
    \fold{9}{9};
  \end{scope}

  \ifheavyonly
  \foreach \y in {0,1.35,2.7,4.05} {
    \begin{scope}[yshift=\y cm]
      \draw[step=1,gray,very thin] (0,0) grid (9,1);
    \end{scope}}
  \fi
}

\def\noiseonobj{
(1.69,0.68),
(2.23,2.84),
(2.52,2.16),
(0.56,3.56),
(0.87,3.19)}

\def\noisesline{
(12.14,.51),
(7.16,.36),
(12.19,.87),
(2.25,.22),
(12.28,.20),
(17.56,.56),
(7.61,.60),
(2.62,.13),
(2.62,.94),
(2.80,.87),
(7.80,.35),
(17.87,.19),
(13.09,.11),
(18.23,.68),
(13.25,.17),
(18.42,.25),
(13.48,.59),
(4.58,.87),
(8.69,.68),
(13.64,.80),
(4.66,.48),
(0.08,.48),
(0.14,.78),
(0.17,.48),
(5.23,.84),
(14.27,.15),
(14.30,.62),
(14.37,.31),
(10.40,.08),
(5.52,.16),
(0.68,.80),
(11.06,.21),
(11.07,.63),
(1.08,.86),
(1.09,.47),
(16.12,.52),
(11.15,.29),
(6.30,.13),
(6.40,.60),
(1.50,.48),
(11.67,.40),
(16.69,.85),
(16.70,.76),
(11.91,.78)}

\def\noisesmod{
0.08/0.48,
0.14/0.78,
0.17/0.48,
0.68/0.80,
1.08/0.86,
1.09/0.47,
1.50/0.48,
2.25/0.22,
2.62/0.13,
2.62/0.94,
2.80/0.87,
4.58/0.87,
4.66/0.48,
5.23/0.84,
5.52/0.16,
6.30/0.13,
6.40/0.60,
7.16/0.36,
7.61/0.60,
7.80/0.35,
8.69/0.68,
10.40/0.08,
11.06/0.21,
11.07/0.63,
11.15/0.29,
11.67/0.40,
11.91/0.78,
12.14/0.51,
12.19/0.87,
12.28/0.20,
13.09/0.11,
13.25/0.17,
13.48/0.59,
13.64/0.80,
14.27/0.15,
14.30/0.62,
14.37/0.31,
16.12/0.52,
16.69/0.85,
16.70/0.76,
17.56/0.56,
17.87/0.19,
18.23/0.68,
18.42/0.25}

\def\noises{
(0.25,0.22),
(0.62,0.13),
(0.62,0.94),
(0.80,0.87),
(1.69,0.68), 
(2.27,0.15),
(2.30,0.62),
(2.37,0.31),
(3.08,0.86),
(3.09,0.47),
(3.50,0.48),
(0.16,1.36),
(0.61,1.60),
(0.80,1.35),
(1.09,1.11),
(1.25,1.17),
(1.48,1.59),
(1.64,1.80),
(2.08,1.48),
(2.14,1.78),
(2.17,1.48),
(2.68,1.80),
(3.30,1.13),
(3.40,1.60),
(0.14,2.51),
(0.19,2.87),
(0.28,2.20),
(1.23,2.68),
(1.42,2.25),
(2.23,2.84),
(2.52,2.16),
(3.06,2.21),
(3.07,2.63),
(3.15,2.29),
(3.67,2.40),
(3.91,2.78),
(0.56,3.56),
(0.87,3.19),
(1.58,3.87),
(1.66,3.48),
(2.40,3.08),
(3.12,3.52),
(3.69,3.85),
(3.70,3.76)}

\ifpageone
\ifmainfile
  \section*{Appendix B\ \ \ The Algorithm in Pictures}
\else
  \clearpage\pagebreak
  \section{The Algorithm in Pictures}
  \label{sec:inpicture}
\fi

\emph{Measurements}\ \ We first compute $Ax'$ from the received signal $x'$. An element of the \mbox{$(N,s,s-r)_q$}-independent code $\mathcal{G}_{CRT} = f_{CRT} \circ \mathcal{H}_P$ is depicted below.
\ifmainfile
  \vspace{.7cm}
\else
  \vspace{15pt}
\fi

\noindent \hspace{-.45cm}
\begin{tikzpicture}[scale=.75]
\begin{scope}
\draw[step=.2,gray,very thin] (0,0) grid (\nn,\nn);
\node[anchor=south] at (2,4.1) {Received signal $x' = x + \mu$};
\node[anchor=north,fill=white] at (2,-.1) {$n$ pixels};
\node[anchor=east,fill=white] at (-.1,2) {$n$};

\node[anchor=west,text width=8cm] at (5,2.5) {There are a total of $N = n^2$ pixels. The goal is to recover the $k$ objects (colored polygons). Each object fits in a $w\times w$ pixel box.};

\begin{scope}[xshift=9.1cm,yshift=.6cm]
  \draw[step=.2,gray,very thin] (0,0) grid (.6,.6);
  \starc{.3}{.23};
  \fill[black!40] (.49,.48) circle(2pt);
  \node[anchor=north] at (.3,-.05) {$w$};
  \node[anchor=east] at (0,.3) {$w$};
\end{scope}

\stara{.4}{3.6};
\starb{2.35}{2.5};
\starc{1.5}{.43};

\foreach \pt in \noises
  \fill[black!40] let \p1 = \pt in (\p1) circle (2pt) ;
\end{scope}

\begin{scope}[yshift=-5.98cm]
\draw[step=1,gray,semithick] (0,0) grid (4,4);
\node[anchor=north,fill=white] at (2,-.1) {$n/w'$ cells};
\node[anchor=east,fill=white] at (-.1,2) {$n/w'$};

\node[anchor=west,text width=12cm] at (5,2.3) {Impose a (randomly shifted) grid $G$ of cells of width \\ $w' = w/\alpha$. For clarity we no longer draw the pixel grid.};

\stara{.4}{3.6};
\starb{2.35}{2.5};
\starc{1.5}{.43};

\foreach \pt in \noises
  \fill[black!40] let \p1 = \pt in (\p1) circle (2pt) ; 
\end{scope}

\begin{scope}[yshift=-9.7cm,scale=.85]
\draw[step=1,gray,semithick] (0,0) grid (19,1);
\stara{17.4}{.6};
\starb{5.35}{.5};
\starc{8.5}{.43};
\foreach \pt in \noisesline
  \fill[black!40] \pt circle (2pt) ; 

\node[anchor=north,fill=white] at (9.5,-.1) {$P > (n/w')^2$ cells};
\node[anchor=east,fill=white] at (-.1,.5) {$1$};

\node[anchor=south,text width=10cm] at (13.8,1.1) {Apply a pairwise independent hash function such as \\ $\mathcal{H}_P : x \rightarrow ax+b\ (\mbox{mod}\ P)$ to a numbering of the cells.};
\end{scope}

\begin{scope}[yshift=-16.95cm,scale=.85]
\measurements;
\node[anchor=south,fill=white] at (4.5,5.2) {Measured signal $z = Ax'$};
\node[anchor=north,fill=white] at (4.5,-.15) {$|B_i| \ge q$ buckets in row $z^i$};
\node[anchor=east,fill=white] at (-.2,2.63) {\large $s$};

\node[anchor=west,text width=5.8cm] at (9.8,2.8) {
Apply an error correcting code $f$ that maps each cell onto exactly one bucket in every row. Sum the cells mapping onto each bucket. The code shown to the left is $f_{CRT}$, where each cell is mapped to its index modulo various (relatively) prime numbers. \\[3pt]
};
\end{scope}
\end{tikzpicture}

\vfill\eject
\fi

\ifpagetwo
\noindent\emph{Recovery}\ \ We now recover from the measurements $Ax'$.
\ifmainfile
  \bigskip
\else
  \vspace{10pt}
\fi

\noindent
\begin{tikzpicture}
\starbwtrue
\def\Axback{.8}
\def\texttab{10}

\begin{scope}[scale=.52]

\begin{scope}[opacity=\Axback]
\measurements
\end{scope}

\node[anchor=west,text width=9cm] at (\texttab,2.1) {Compute the feature vector $F(z_j^i)$ of each bucket. \\ In our experiments, the feature vector contains \\ information about mass and centroid.};
\node (A) at (5.5,.2) {};
\node (A1) at (6.4,-.85) {};
\draw[thick] (5,0) rectangle (6,1);
\draw[thick] (A1) node[right,text width=3cm]
        {$\mathsf{F(z_6^4) = (8, 15, 9.2)}$} to [bend left] (A);

\node (B) at (4.8,4.55) {};
\node (B1) at (6.5,5.65) {};
\draw[thick] (4,4.05) rectangle (5,5.05);
\draw[thick] (B1) node[right, text width=3cm]
        {$\mathsf{F(z_5^1) = (2, 3, 1.1)}$} to [out=180,in=0] (B);

\node (C) at (5.8,3.2) {};
\node (C1) at (7.5,4.45) {};
\draw[thick] (5,2.7) rectangle (6,3.7);
\draw[thick] (C1) node[right, text width=3cm]
        {$\mathsf{F(z_6^2) = (8, 14, 9.5)}$} to [out=180,in=0] (C);

\begin{scope}[yshift=-7.8cm]
\begin{scope}[opacity=\Axback]
\heavyonlytrue
\measurements
\end{scope}
\node[anchor=west,text width=9cm] at (\texttab,3.3) {Set $R =\{ (i,j) : \|F(z_j^i)\|_\Gamma$ is large$\}$. \\Discard buckets not in $R$.};

\foreach \x/\y in {2/0,4/0,5/0,8/0, 0/1,1/1,2/1,5/1, 0/2,1/2,3/2,4/2,5/2, 1/3,2/3,3/3}
  \draw[very thick, brown] (\x,\y*1.35) rectangle (\x+1,\y*1.35+1);

\begin{scope}[xshift=6cm,yshift=1.2cm]
\node[anchor=east] at (\texttab,0) {$R =\ $};
\draw[very thick,brown] (\texttab,-.55) rectangle (\texttab+1,.45);
\end{scope}
\end{scope} 

\begin{scope}[yshift=-15cm]
\begin{scope}[opacity=\Axback]
\heavyonlytrue
\measurements
\end{scope}
\node[anchor=west,text width=9cm] at (\texttab,3.2) {Cluster $F(R) =\{ F(z_j^i) : (i,j) \in R\}$ into $k$ clusters\\ (with outliers). This induces a partition $R',R^1 \ldots R^k$\\ of $R$, with $F(R^l)$ equal to the $l$-th cluster.};

\foreach \x/\y in {1/3,3/2,1/1}
  \draw[r1] (\x,\y*1.35) rectangle (\x+1,\y*1.35+1);
\foreach \x/\y in {3/3,1/2,0/1,2/0}
  \draw[r2] (\x,\y*1.35) rectangle (\x+1,\y*1.35+1);
\foreach \x/\y in {5/2,5/1,5/0}
  \draw[r3] (\x,\y*1.35) rectangle (\x+1,\y*1.35+1);

\begin{scope}[xshift=3.4cm,yshift=.7cm]
\foreach \s/\x/\i in {r1/0/1, r2/3.3/2, r3/6.6/3} {
  \node[anchor=east] at (\x+\texttab,0) {$R^\i =\ $};
  \draw[\s] (\x+\texttab,-.55) rectangle (\x+\texttab+1,.45);
}
\end{scope}

\end{scope} 

\begin{scope}[yshift=-22.2cm]

\begin{scope}[xshift=1.7cm,scale=1.25]
\draw[step=1,gray,very thin] (0,0) grid (4,4);
\draw[gray,thin] (0,0) rectangle (4,4);
\draw[r1] (0,3) rectangle (1,4);
\draw[r2] (1,0) rectangle (2,1);
\draw[r3] (2,2) rectangle (3,3);
\end{scope}

\node[anchor=west,text width=11cm] at (\texttab-1.8,2.65) {Decode each $R^l$ to obtain a cell $d^l$ in the original image.};
\end{scope} 

\begin{scope}[yshift=-29.2cm]
\begin{scope}[xshift=1.7cm,scale=1.25]
\draw[step=1,gray,very thin] (0,0) grid (4,4);
\draw[gray,thin] (0,0) rectangle (4,4);
\draw[r1] (0,3) rectangle (1,4);
\draw[r2] (1,0) rectangle (2,1);
\draw[r3] (2,2) rectangle (3,3);
\starbwfalse
\stara{.4}{3.6};
\starb{2.35}{2.5};
\starc{1.5}{.43};
\foreach \pt in \noiseonobj
  \fill[black!40] \pt circle (2pt) ;
\end{scope}

\node[anchor=west,text width=9cm] at (\texttab-1.8,2.7) {Though we don't elaborate in the text, a simple min or median process can be used to obtain an approximation for the contents of each $d^l$.};
\end{scope} 

\end{scope}
\end{tikzpicture}
\fi

\ifmainfile
  \end{document}
\else
\fi

\hypersetup{urlcolor=black}
\bibliographystyle{alpha}
\bibliography{main-arxiv}

\newcommand{\etalchar}[1]{$^{#1}$}
\begin{thebibliography}{CKMN01}

\bibitem[{Bal}]{Ball}
{Ball Aerospace \& Technologies Corp}.
\newblock {CT-602} star tracker.
\newblock Linked from \url{http://www.ballaerospace.com/page.jsp?page=104}.

\bibitem[BGI{\etalchar{+}}08]{BGIKS08}
R.~Berinde, A.C. Gilbert, P.~Indyk, H.~Karloff, and M.J. Strauss.
\newblock Combining geometry and combinatorics: A unified approach to sparse
  signal recovery.
\newblock In {\em 46th Annual Allerton Conf. on Communication, Control, and
  Computing}, pages 798--805. IEEE, 2008.

\bibitem[BI09]{BI09}
R.~Berinde and P.~Indyk.
\newblock Sequential sparse matching pursuit.
\newblock In {\em 47th Annual Allerton Conf. on Communication, Control, and
  Computing}, pages 36--43. IEEE, 2009.

\bibitem[CCF04]{CCF}
M.~Charikar, K.~Chen, and M.~{Farach-Colton}.
\newblock Finding frequent items in data streams.
\newblock {\em Theoretical Comp. Sci.}, 312(1):3--15, 2004.

\bibitem[CKMN01]{CKMN}
M.~Charikar, S.~Khuller, D.M. Mount, and G.~Narasimhan.
\newblock Algorithms for facility location problems with outliers.
\newblock In {\em Proc. 12th Annual Symp. on Discrete Algorithms}, pages
  642--651. SIAM, 2001.

\bibitem[CM05]{CM03b}
G.~Cormode and S.~Muthukrishnan.
\newblock An improved data stream summary: the count-min sketch and its
  applications.
\newblock {\em J. Algorithms}, 55(1):58--75, 2005.

\bibitem[{Com}]{Comtech}
{Comtech AeroAstro}.
\newblock Miniature star tracker data sheet.
\newblock Linked from
  \url{http://www.aeroastro.com/index.php/space-products-2/minature-star-track%
er-mst}.

\bibitem[CRT06]{CRT06:Stable-Signal}
E.~J. Cand{\`e}s, J.~Romberg, and T.~Tao.
\newblock Stable signal recovery from incomplete and inaccurate measurements.
\newblock {\em Comm.\ on Pure and Applied Math.}, 59(8):1208--1223, 2006.

\bibitem[DDT{\etalchar{+}}08]{DDTLTKB}
M.F. Duarte, M.A. Davenport, D.~Takhar, J.N. Laska, T.~Sun, K.F. Kelly, and
  R.G. Baraniuk.
\newblock Single-pixel imaging via compressive sampling.
\newblock {\em Signal Processing Magazine, IEEE}, 25(2):83--91, 2008.

\bibitem[DIPW10]{DIPW}
K.~{Do Ba}, P.~Indyk, E.~Price, and D.P. Woodruff.
\newblock Lower bounds for sparse recovery.
\newblock In {\em Proc. 21st Annual Symp. on Discrete Algorithms}, pages
  1190--1197. SIAM, 2010.

\bibitem[Don06]{Don06:Compressed-Sensing}
D.~L. Donoho.
\newblock {C}ompressed {S}ensing.
\newblock {\em IEEE Trans. Info. Theory}, 52(4):1289--1306, 2006.

\bibitem[Fos93]{Fos}
E.~R. Fossum.
\newblock {A}ctive {P}ixel {S}ensors: Are {CCD}'s dinosaurs?
\newblock {\em SPIE}, 1900:2--14, 1993.

\bibitem[FPRU10]{FPRU}
S.~Foucart, A.~Pajor, H.~Rauhut, and T.~Ullrich.
\newblock The gelfand widths of lp-balls for $0< p <= 1$.
\newblock {\em J. Complexity}, 26(6):629--640, 2010.

\bibitem[FTF06]{FTF}
R.~Fergus, A.~Torralba, and W.~T. Freeman.
\newblock Random lens imaging.
\newblock {\em MIT CSAIL Technical Report 2006-058}, 2006.

\bibitem[GI10]{GI}
A.~Gilbert and P.~Indyk.
\newblock Sparse recovery using sparse matrices.
\newblock {\em Proc. IEEE}, 98(6):937--947, 2010.

\bibitem[GRS00]{GRS99}
O.~Goldreich, D.~Ron, and M.~Sudan.
\newblock Chinese remaindering with errors.
\newblock {\em IEEE Trans. Information Theory}, 46(4):1330--1338, 2000.

\bibitem[Gur10]{Guru}
V.~Guruswami.
\newblock Introduction to coding theory.
\newblock {\em Course notes}, Lecture 1, 2010.
\newblock Available at
  \url{http://www.cs.cmu.edu/~venkatg/teaching/codingtheory/notes/notes1.pdf}.

\bibitem[HL07]{Holst07}
G.~C. Holst and T.~S. Lomheim.
\newblock {\em CMOS/CCD Sensors and Camera Systems}.
\newblock JCD Publishing and SPIE Press, 2007.

\bibitem[Hol98]{Hol}
G.~C. Holst.
\newblock {\em {CCD} Arrays, Cameras, and Displays}, pages 79--82, 91--93,
  123--127.
\newblock {JCD} Publishing and SPIE Optical Engineering Press, second edition,
  1998.

\bibitem[Ind07]{I-SSS}
P.~Indyk.
\newblock Sketching, streaming and sublinear-space algorithms.
\newblock {\em Graduate course notes}, 2007.
\newblock Available at \url{http://stellar.mit.edu/S/course/6/fa07/6.895/}.

\bibitem[Jus76]{Jus76}
J.~Justesen.
\newblock {On the complexity of decoding Reed-Solomon codes (Corresp.)}.
\newblock {\em IEEE Trans. Information Theory}, 22(2):237--238, 1976.

\bibitem[Lie02]{Lie}
C.~C. Liebe.
\newblock Accuracy performance of star trackers --- a tutorial.
\newblock {\em IEEE Trans. Aerospace and Electronic Systems}, 38(2):587--599,
  2002.

\bibitem[Lit01]{Litw}
Dave Litwiller.
\newblock {CMOS} vs. {CCD}: Facts and fiction.
\newblock {\em Photonics Spectra}, January 2001.

\bibitem[LOA05]{LOA}
S.~Lee, G.~G. Ortiz, and J.~W. Alexander.
\newblock Star tracker-based acquisition, tracking, and pointing technology for
  deep-space optical communications.
\newblock {\em Interplanetary Network Progress Report}, (42-161), 2005.

\bibitem[Mei03]{Meindl03}
James~D. Meindl.
\newblock Beyond moore's law: The interconnect era.
\newblock {\em Computing in Science and Engineering}, 5:20--24, 2003.

\bibitem[MJS{\etalchar{+}}10]{Majidzadeh10}
V~Majidzadeh, L~Jacques, A~Schmid, P~Vandergheynst, and Y~Leblebici.
\newblock A (256x256) {P}ixel 76.7{m}{W} {CMOS} {I}mager/{C}ompressor {B}ased
  on {R}eal-{T}ime {I}n-{P}ixel {C}ompressive {S}ensing.
\newblock In {\em {IEEE} {I}nt. {S}ymp. {C}ircuits and {S}ystems}, 2010.

\bibitem[Mor97]{Mor}
D.~Mortari.
\newblock Search-less algorithm for star pattern recognition.
\newblock {\em J. Astronautical Sciences}, 45(2):179--194, 1997.

\bibitem[Mut05]{Muthu:survey}
S.~Muthukrishnan.
\newblock Data streams: Algorithms and applications.
\newblock {\em Foundations and Trends in Theoretical Comp. Sci.}, 2005.

\bibitem[Nac10]{Me}
M.~Nachin.
\newblock Lower bounds on the column sparsity of sparse recovery matrices.
\newblock {\em MIT Undergraduate Thesis}, 2010.

\bibitem[RGC{\etalchar{+}}10]{Robucci10}
R.~Robucci, J.D. Gray, L.K. Chiu, J.~Romberg, and P.~Hasler.
\newblock Compressive sensing on a {CMOS} separable-transform image sensor.
\newblock {\em Proc. IEEE}, 98(6):1089--1101, 2010.

\bibitem[Rom09]{Rom}
J.~Romberg.
\newblock Compressive sampling by random convolution.
\newblock {\em SIIMS}, 2009.

\bibitem[SM09]{SM}
B.~B. {Spratling IV} and D.~Mortari.
\newblock A survey on star identification algorithms.
\newblock {\em Algorithms}, 2:93--107, 2009.

\bibitem[Smi]{SAO}
Smithsonian astrophysical observatory star catalog.
\newblock Available at
  \url{http://heasarc.gsfc.nasa.gov/W3Browse/star-catalog/sao.html}.

\bibitem[TAN10]{TAN}
V.~Treeaporn, A.~Ashok, and M.~A. Neifeld.
\newblock Increased field of view through optical multiplexing.
\newblock {\em Optics Express}, 18(21), 2010.

\bibitem[UGN{\etalchar{+}}09]{UGNKJKB}
S.~Uttam, A.~Goodman, M.~Neifeld, C.~Kim, R.~John, J.~Kim, and D.~Brady.
\newblock Optically multiplexed imaging with superposition space tracking.
\newblock {\em Optics Express}, 17(3), 2009.

\bibitem[WL99]{RL}
J.~R. Wertz and W.~J. Larson.
\newblock {\em Space mission analysis and design}, pages 315--316, 321--322,
  894--895.
\newblock Space technology series. Microcosm press and Kluwer academic
  publishers, third edition, 1999.

\end{thebibliography}

\end{document}